\documentclass[openacc]{rsproca_new}
\usepackage{float}
\usepackage{hyperref}
\usepackage[labelfont=bf]{caption}
\usepackage{threeparttable}
\usepackage{subcaption}
\usepackage{xcolor}
\usepackage[export]{adjustbox}
\usepackage[bottom]{footmisc}
\usepackage{bm}

\usepackage{amsmath,amsthm,amssymb,commath}
\usepackage{accents}
\newlength{\dhatheight}

\usepackage[mathscr]{euscript}
\usepackage{textcomp}
\usepackage{graphicx,psfrag}

\newcommand{\tx}{\text}
\newcommand{\scr}{\mathscr}
\newcommand{\be}{\begin{equation}}
\newcommand{\en}{\end{equation}}
\newcommand{\la}{\label}

\def\rr#1{(\ref{#1})}
\renewcommand{\vec}[1]{\mathbf{#1}}

\begin{document}

\title{Post-bifurcation behaviour of elasto-capillary necking and bulging in soft tubes}

\author{
Dominic Emery$^{1}$ and Yibin Fu$^{1}$}

\address{$^{1}$ School of Computing and Mathematics, Keele University, Staffordshire ST5 5BG, U.K.}

\subject{Applied mathematics}


\corres{Dominic Emery\\
\email{d.r.emery@keele.ac.uk}}

\begin{abstract}
Previous linear bifurcation analyses have evidenced that an axially stretched soft cylindrical tube may develop an infinite-wavelength (localised) instability when one or both of its lateral surfaces are under sufficient surface tension. Phase transition interpretations have also highlighted that the tube admits a final evolved "two-phase" state. How the localised instability initiates and evolves into the final "two-phase" state is still a matter of contention, and this is the focus of the current study.
Through a weakly non-linear analysis conducted for a general material model, the initial \textit{sub-critical} bifurcation solution is found to be localised bulging or necking depending on whether the axial stretch is greater or less than a certain threshold value. At this threshold value, an exceptionally \textit{super-critical} kink-wave solution arises in place of localisation. A thorough interpretation of the anticipated post-bifurcation behaviour based on our theoretical results is also given, and this is supported by Finite Element Method (FEM) simulations.
\end{abstract}
\begin{fmtext}
\section{Introduction \label{sec1}}
A surge of interest in the behaviour, functionality and development of micro and nano-scale soft materials has transpired in recent years, with applications in soft robotics \cite{wang2018soft} and the construction of artificial muscles \cite{qiu2019} and other biomedical devices \cite{cooke2018} being at the forefront of this new-found motivation. A bi-product of this is the emerging field of \textit{elasto-capillarity}, which is concerned with the finite deformation of elastic solids with surface energy \cite{saksono2006,javili2009,lf2012,papastavrou2013,liu2017closed,bico2018}. This surface energy becomes non-negligible when the typical length scale of a system is comparable to the ratio of the surface tension $\gamma$ to the ground state shear modulus $\mu$ \cite{style}. Thus, when modelling extremely soft materials such as gels, elastomers and biological tissue on the nano to milli-scale, the consideration of elasto-
\end{fmtext}
\maketitle
\noindent capillary effects is of vital importance. A geometry which arguably requires greater attention is the cylindrical tube, which is widespread in physiological systems in the form of arteries, airways and intestines, for instance. The villification of the gastrointestinal tract \cite{shyer2013}, the closure of pulminary airways \cite{seow2000} and the gyrification of the brain \cite{balbi2020} are examples of physiological tubular instabilities which have predominantly been treated as purely growth induced, with little attention given to elasto-capillary effects. Exceptionally, consideration is given to the surface tension induced buckling of liquid lined tubes as a model for airway closure in \cite{hazel2005}, and insights into elasto-capillary circumferential buckling instabilities in tubes under axial loading \cite{emery2021elasto}, growth \cite{riccobelli2020} and uniform pressure and geometric everting \cite{wang2021large} have very recently transpired.

\begin{figure}[h!]
\vspace{2mm}
\centering
\includegraphics[scale=0.3]{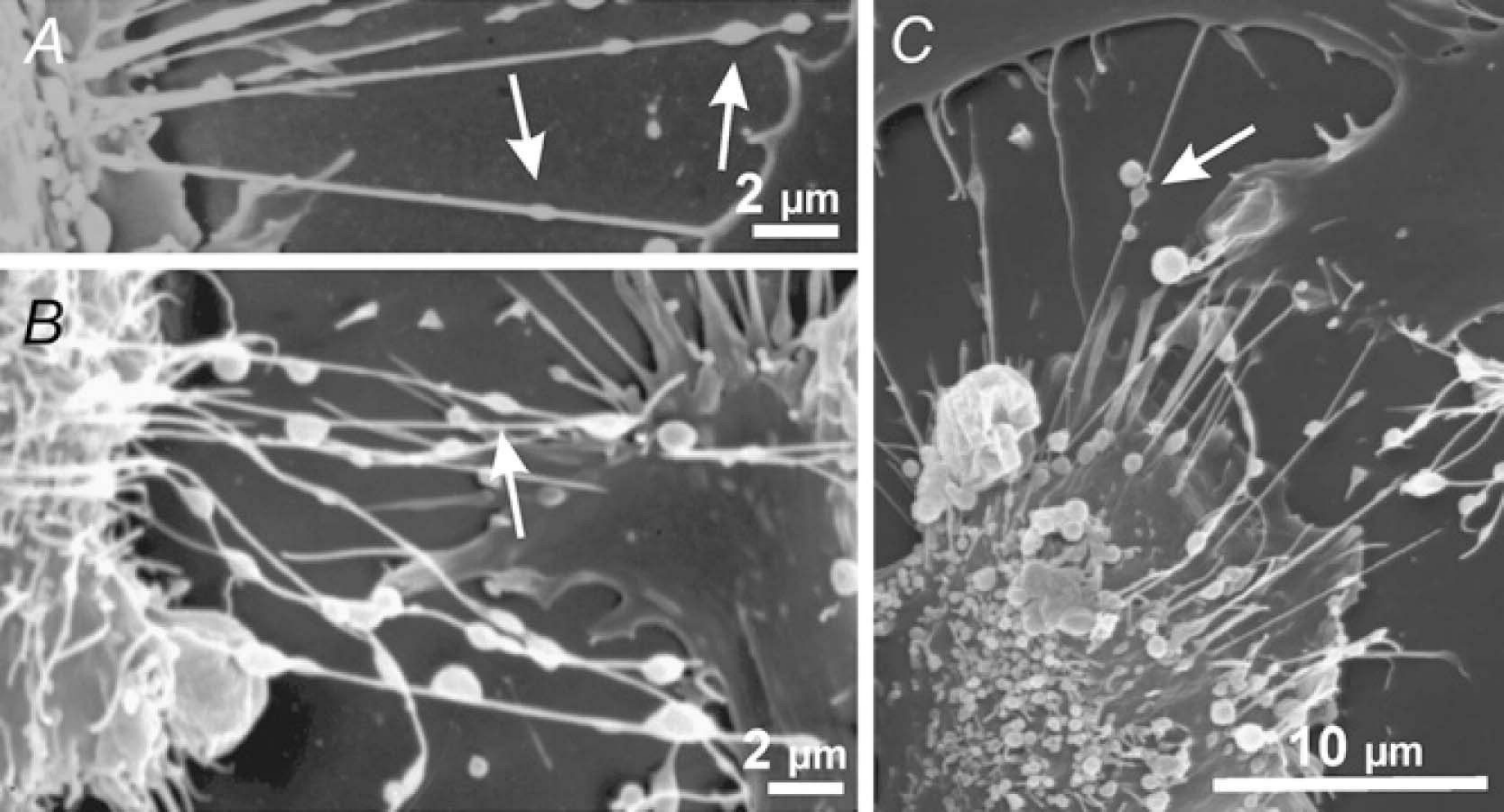}
\vspace{2mm}
\caption{Tunnelling nanotubes (TNT's) have been observed between cells which migrate from one another, and are subject to axial stretching as the cells move further apart. Such tubes have been found to develop static localised axisymmetric beads\cite{ver2008}.}
\label{fig1}
\end{figure}

The well known peristaltic instability in which soft slender cylinders/tubes adopt axisymmetric beads under bulk and surface stresses has drawn much attention in recent years\cite{barriere1996,mora2010,ciarletta2012peristaltic,taffetani,xuan2016,lestringant2020}. This bifurcation phenomenon can occur in stretched nerve fibres \cite{bar1994}, in axons under mechanical trauma \cite{kilinc,goriely2015}, and it has also been implicated in neurodegenerative disorders such as Alzheimer's and Parkinson's diseases \cite{datar2019}. Also, stretched tubes termed \textit{tunnelling nanotubes} have been observed between migrating cells, and allow for inter-cellular communications and migration support \cite{rustom2004,schara2009,sis2015,vig2017,drab2019,ali2020}. As is shown in Fig. $\ref{fig1}$, the formation of static localised beads has in-fact been observed in these nanotubes \cite{ver2008}. Theoretically, it has been evidenced in the case of a solid cylinder that the beading instability culminates in a "two-phase" state, or a "kink-wave" solution, characterised by two sections with distinct but uniform axial stretch connected by a smooth transition zone \cite{xuan2017,giudici2020}. Moreover, it was determined in \cite{FuST} that a localised bulging or necking solution will initially occur depending on the loading path.

The beading instability is also observable in hollow tubes which are filled with magnetic fluids \cite{menager2002}, submerged in hydrophillic polymer solutions \cite{tsafrir2001} and under growth \cite{hannezo2012}, and has also been implicated in the synthesis of soft matter nanotubes \cite{ma2017} which have a variety of physical, biological and chemical applications \cite{shimizu2020}.


In spite of this, only very recently have the first steps been taken towards obtaining a concrete theoretical understanding of beading in hollow tubes. An infinite-wavelength instability in a cylindrical cavity under surface tension was first found to be preferred in \cite{xuan2016}. Then, a linear bifurcation analysis of tubes under surface tension and axial stretching was initially conducted by \cite{liuwang}, with further insights provided by \cite{EmeryFu2020}. In the latter, necessary conditions for localised bifurcation are determined for three distinct boundary conditions. In Case 1, both lateral boundaries are traction-free and under surface tension, whilst in Case 2 (resp. Case 3), the inner (resp. outer) lateral surface is in smooth contact with a rigid boundary to prevent radial displacement and surface tension (with the other surface remaining traction-free and under surface tension). Cases 2 and 3 have previously been investigated through FEM simulations \cite{henann}, with motivation stemming from the fact that the two types of boundary conditions seem to appear in
many biological systems. Indeed, consideration of these different boundary conditions allows
us to analyse how the localised instability is affected by different constraints. For instance, a change in results will later be highlighted when the (scaled) initial inner radius A < 0.08567 in Case 3.

In Case 1, we originally thought in \cite{EmeryFu2020} that localised bifurcation was associated with negative surface tension, which is physically implausible. However, we have now discovered that localised solutions can in theory exist in Case 1, but are less favourable than circumferential buckling modes \cite{emery2021elasto}; an explanation of this change in viewpoint is given shortly. In Cases 2 and 3, the linear bifurcation analysis in \cite{EmeryFu2020} showed that localisation is favourable over periodic axisymmetric modes. To provide further understanding of the beading instability in hollow tubes, we extend this work here by conducting a weakly non-linear near-critical analysis for Cases 2 and 3. We show that a localised bifurcation solution generally initiates sub-critically, and that the explicit nature of this solution is highly dependant on the value of the principal axial stretch $\lambda$ and the loading path. Through this theoretical analysis, we are able to give in-depth insights into the post-bifurcation behaviour for multiple loading scenarios. Namely, we interpret the evolution from the initial localised solution to a final "two-phase" state, and this is supported by previous FEM simulations \cite{EmeryFu2020}.


We remark that the localised solutions discussed in the current study are essentially solitary waves with zero wave speed.  The necking and bulging solutions correspond to static "dark" and "bright" solitons, respectively. Solitary waves were first observed in the context of water waves by Russell \cite{russell1845}, and the associated model equation was first derived by Korteweg and De Vries \cite{korteweg1895} and is nowadays known as the KdV equation. The other simplest model equation that admits a solitary wave solution is the nonlinear Schr\"odinger equation (NLSE) which was first derived in \cite{chiao1965} for propagation of light in nonlinear optical fibers (mathematically the amplitude evolution of wave trains). The static counterpart of NLSE has been derived to describe the amplitude variation of periodic buckling modes \cite{lange1971,potier1987}. In recent decades, buckling of an Euler beam on a nonlinear foundation has been much studied in relation to localised solutions \cite{hunt2000}. Such localised solutions again correspond to amplitude localisation of periodic buckling modes. A huge variety of other model equations have also been derived for a range of physical processes to incorporate additional effects and/or to describe degenerate cases. Some of these equations involve higher order spatial derivatives and multi spatial dimensions, e.g. the Swift-Hohenberg equation for thermal convection \cite{swift1977}. We refer to the monograph by Peletier and Troy \cite{peletier2001} for a discussion of some of these equations. Physically speaking, solitary waves arise from a balance of nonlinearity and dispersion, and this balance underpins all the amplitude equations that admit solitary wave solutions.  For a dynamical systems theory point of view, we refer to one of the earliest papers by Kirchg\"assner \cite{kirch} and the more recent monograph by Haragus and Iooss \cite{Iooss2010}.

The remainder of this paper is organised as follows. After formulating the problem in the next section, we extend in section $\ref{sec3}$ the derivation of the primary deformation and analytical bifurcation conditions for localisation (presented originally in \cite{EmeryFu2020} for the neo-Hookean strain-energy function) to a general material model. We also explain why, contrary to our original claims, localised bifurcation is \textit{theoretically} possible in Case 1. We then conduct a thorough weakly non-linear analysis in section $\ref{sec4}$ for Case 2 to decipher whether a localised solution can actually bifurcate from the primary state, and also to determine the explicit nature of possible bifurcations. For multiple loading paths, a complete interpretation of the anticipated post-bifurcation behaviour is also given. Then, in section $\ref{sec6}$, we comment on the distinctions in the corresponding analysis for Case 3, focussing primarily on the regime of large thickness. Finally, concluding remarks are given in section $\ref{sec7}$. All of our computations and algebraic manipulations were performed in \textit{Mathematica} \cite{wo2019}, and the supplementary code is available on request.

\section{Problem formulation \label{sec2}}

Consider a hyperelastic cylindrical tube with an initial inner radius $A$, outer radius $B$ and axial half-length $L\gg B$. We use cylindrical polar coordinates
$(R, \Theta, Z)$ and $(r, \theta, z)$ to describe the position vectors $\vec{X}$ and $\vec{x}$ of a representative material particle in the undeformed and deformed configurations, respectively.
Under the general axisymmetric deformation
\begin{align}
r&=r\left(R,Z\right),\,\,\,\,\,\,\,\,\theta=\Theta,\,\,\,\,\,\,\,\,z=z\left(R,Z\right), \label{gendef}
\end{align}
the inner and outer radii become $a$ and $b$, respectively, whilst the axial half-length becomes $\ell\gg b$. The deformation gradient $\vec{F}$ is then defined through $d\vec{x}=\vec{F}\,d\vec{X}$ and takes the form
\begin{align}
\vec{F}&=\frac{\partial r}{\partial R}\,\vec{e}_r\otimes\vec{E}_R+\frac{\partial r}{\partial Z}\,\vec{e}_r\otimes\vec{E}_Z+\frac{r}{R}\,\vec{e}_\theta\otimes\vec{E}_{\Theta}+\frac{\partial z}{\partial R}\,\vec{e}_z\otimes\vec{E}_R+\frac{\partial z}{\partial Z}\,\vec{e}_z\otimes\vec{E}_Z, \label{Fgen}
\end{align}
where $(\vec{E}_{R},\,\vec{E}_{\Theta},\,\vec{E}_{Z})$ and $(\vec{e}_{r},\,\vec{e}_{\theta},\,\vec{e}_{z})$ are the orthonormal bases associated with the previously defined sets of coordinates. Assuming that the tube material is incompressible, we enforce the following constraint of isochorism:
\begin{align}
\text{det}\,\vec{F}&=1. \label{detF}
\end{align}
A strain-energy function $w$, which governs the constitutive behaviour of the tube, can then be introduced as follows:
\begin{align}
w&=w\left(I_B\right), \label{SEfunction}
\end{align}
where $I_B$ is the first principal invariant of the left Cauchy-Green strain tensor $\vec{B}=\vec{F}\vec{F}^T$, i.e. $I_B=\text{tr}\,\vec{B}$, with the superscript $T$ denoting transposition. This class of strain-energy functions has been shown to be suitable for many different materials under tension \cite{wineman2005}.
 Two of the most common strain-energy functions of this form are the neo-Hookean and Gent material models, which take the respective forms
\begin{equation}
w=\frac{1}{2}\,\mu\left(I_B -3\right)\,\,\,\,\,\,\,\,\text{and}\,\,\,\,\,\,\,\,w=-\frac{1}{2}\mu\,J_{\tx{m}}\ln\left(1-\frac{I_{B}-3}{J_{\tx{m}}}\right), \label{neohook}
\end{equation}
where $J_{\tx{m}}$ is a constant governing the extensibility limit of the material. We note that, in the limit $J_{\tx{m}}\rightarrow\infty$, the neo-Hookean material model is recovered from $(\ref{neohook})_{2}$. For the remainder of this paper we scale all lengths by $B$, all stresses by $\mu$ and the surface tension $\gamma$ by $\mu B$. Therefore, we may set $B=1$ and $\mu=1$ without loss of generality.

\subsection{Stream-function formulation}
 The problem can be elegantly re-formulated in terms of a single mixed co-ordinate stream function $\phi=\phi\,(R,\,z)$ so that the incompressibility constraint $(\ref{detF})$ is satisfied exactly \cite{ciarletta2011}. This stream function is defined through the relations
\begin{align}
 r^2&=2\,\phi_{,z},\,\,\,\,\,\,\,\,Z=\frac{1}{R}\,\phi_{,R}, \label{incphi}
\end{align}
where a comma denotes partial differentiation with respect to the implied coordinate. Accordingly, $\vec{F}$ can be re-written in the form
\begin{align}
\vec{F}&=\frac{1}{\sqrt{2\,\phi_{,z}}}\left[\phi_{,Rz}-R\,\frac{\phi_{,zz}}{\phi_{,Rz}}\frac{\partial}{\partial R}\left(\frac{1}{R}\phi_{,R}\right)\right]\,\vec{e}_r\otimes\vec{E}_R + \frac{R\,\phi_{,zz}}{\sqrt{2\,\phi_{,z}}\,\phi_{,Rz}}\,\vec{e}_r\otimes\vec{E}_Z + \frac{\sqrt{2\,\phi_{,z}}}{R}\,\vec{e}_{\theta}\otimes\vec{E}_{\Theta} \nonumber\\[1em]
&\,\,\,\,\,\,\,-\frac{R}{\phi_{,Rz}}\frac{\partial}{\partial R}\left(\frac{1}{R}\phi_{,R}\right)\,\vec{e}_z\otimes\vec{E}_R + \frac{R}{\phi_{,Rz}}\,\vec{e}_z\otimes\vec{E}_Z, \label{Fphi}
\end{align}
and the invariant $I_{B}$ may then be computed from $(\ref{Fphi})$. In Case 1 where both lateral boundaries are under surface tension, the total energy $\mathcal{E}$ of the static axisymmetric solution is the sum of the bulk elastic energy and the surface energies on these boundaries, i.e.:
\begin{align}
\mathcal{E}&=2\,\pi\int^{\ell}_{-\ell}\int^{B}_{A}\,\mathcal{L}_{b}\,dR\,dz \,+\, 2\,\pi\int^{\ell}_{-\ell}\,\left(\mathcal{L}_{s}^{A}+\mathcal{L}_{s}^{B}\right)\,dz, \label{E}
\end{align}
where the bulk Lagrangian $\mathcal{L}_{b}$ and surface Lagrangians $\mathcal{L}_{s}^{A,B}$ are defined through
\begin{align}
\mathcal{L}_b&=\phi_{,Rz}\,w(I_B),\,\,\,\,\,\,\,\,\mathcal{L}_s^{A,B}=\gamma\left.\sqrt{2\,\phi_{,z}+\phi_{,zz}^2}\right\vert_{R=A,B}.
\end{align}
Equilibrium of bulk elastic forces requires we satisfy Euler-Lagrange equation given by
\begin{align}
\left(\frac{\partial \mathcal{L}_b}{\partial \phi_{,\beta\Gamma}}\right)_{,\beta\Gamma}-\left(\frac{\partial \mathcal{L}_b}{\partial \phi_{,\delta}}\right)_{,\delta}&=0. \label{goveqn}
\end{align}
The standard summation convention is applied here, with $\delta=R$ or $z$ and $\beta\Gamma = RR$, $Rz$ or $zz$. The normal traction-free boundary conditions on the inner and outer lateral surfaces take the respective forms
\begin{align}
\frac{\partial \mathcal{L}_b}{\partial \phi_{,R}}-\left(\frac{\partial \mathcal{L}_b}{\partial \phi_{,RR}}\right)_{,R}-\left(\frac{\partial \mathcal{L}_b}{\partial \phi_{,Rz}}\right)_{,z}-\left(\frac{\partial \mathcal{L}_s^{A}}{\partial \phi_{,zz}}\right)_{,zz}+\left(\frac{\partial \mathcal{L}_s^{A}}{\partial \phi_{,z}}\right)_{,z}&=0,\,\,\,\,\,\,\,\,\,\,\,\,\,\,R=A, \label{BC1A} \\[1em]
\frac{\partial \mathcal{L}_b}{\partial \phi_{,R}}-\left(\frac{\partial \mathcal{L}_b}{\partial \phi_{,RR}}\right)_{,R}-\left(\frac{\partial \mathcal{L}_b}{\partial \phi_{,Rz}}\right)_{,z}+\left(\frac{\partial \mathcal{L}_s^{B}}{\partial \phi_{,zz}}\right)_{,zz}-\left(\frac{\partial \mathcal{L}_s^{B}}{\partial \phi_{,z}}\right)_{,z}&=0,\,\,\,\,\,\,\,\,\,\,\,\,\,\,R=B,\label{BC1B}
\end{align}
where the change in sign of the surface energy terms is due to the opposing mean curvatures of the inner and outer boundaries. Where a lateral surface is in smooth contact with a rigid boundary (i.e. Cases 2 and 3 detailed previously), the zero normal traction condition on this surface is replaced by the requirement that the radial displacement vanishes. This condition will be introduced at a later stage in this paper. In all three cases, we have zero shear traction on both lateral surfaces, invoking a further two boundary conditions which are expressed as follows:
\begin{align}
\frac{\partial \mathcal{L}_b}{\partial \phi_{,RR}}&=0,\,\,\,\,\,\,\,\,\,\,\,\,\,\,R=A,\,B.  \label{BC2}
\end{align}

\section{Primary deformation and bifurcation conditions for localisation \label{sec3}}

We now narrow our focus towards the following primary axi-symmetric deformation, a sub-class of $(\ref{gendef})$,  which is theoretically possible for all strain-energy functions:
\begin{align}
r=r(R),\,\,\,\,\,\,\,\,\theta=\Theta,\,\,\,\,\,\,\,\,z=\lambda Z, \label{axdef}
\end{align}
with $\lambda$ defined as the principal axial stretch. The deformation gradient corresponding to $(\ref{axdef})$ is
\begin{align}
\vec{F}&=\frac{\partial r}{\partial R}\,\vec{e}_r\otimes\vec{E}_R + \frac{r}{R}\,\vec{e}_{\theta}\otimes\vec{E}_{\Theta}+\lambda\,\vec{e}_z\otimes\vec{E}_Z. \label{Fax}
\end{align}
We consider Cases 1, 2 and 3  separately for the remainder of this section.

\subsection{Case 1 - Traction-free lateral boundaries under surface tension \label{sec3a}}

In Case 1, upon substitution of $(\ref{Fax})$ into $(\ref{detF})$, the primary radial displacement $r_{0}$ which satisfies incompressibility exactly is found to take the form
\begin{align}
r_{0}&=\sqrt{\lambda^{-1}\left(R^2-A^2\right)+a^2}, \label{rR}
\end{align}
with the outer deformed radius $b$ then becoming $r_{0}(B)$ and with $a$ being an unknown parameter. Through integration of $(\ref{incphi})$, the primary solution for $\phi$, denoted $\phi_0$, and the associated expression for $I_{B}$ are given by
\begin{align}
\phi_0&=\frac{R^2 z}{2\,\lambda}+\frac{1}{2}\left(a^{2}-\frac{A^{2}}{\lambda}\right)z\,\,\,\,\,\,\,\,\text{and}\,\,\,\,\,\,\,\,I_{0}(R)=\left.I_{B}\right\vert_{\phi_{0}}=\frac{\left(A^{2}-a^{2}\lambda\right)^{2}}{r_{0}^{2}\,R^{2}\,\lambda^{2}}+\frac{2+\lambda^{3}}{\lambda}. \label{phi0gen}
\end{align}
The tube is under the combined action of surface tension $\gamma$ on $R=A,\,B$  and a resultant axial force $\mathcal{N}$, such that the total energy of the primary state is
\begin{align}
\mathcal{E}&=2\,\pi\left[\,\int^{\lambda L}_{-\lambda L}\int^{B}_{A}\,\mathcal{L}_{b}\,dR\,dz \,+\,\int^{\lambda L}_{-\lambda L}\,\left(\mathcal{L}_{s}^{A}+\mathcal{L}_{s}^{B}\right)\,dz\,\right] - (\lambda -1)\,\mathcal{N},\,\,\,\,\,\,\,\,\text{with}\,\,\,\,\,\,\,\,\phi=\phi_{0}. \label{TPEcase1}
\end{align}
Equilibrium of the primary state requires that we satisfy $\partial\mathcal{E}/\partial \lambda =0$ and $\partial\mathcal{E}/\partial a=0$, and from these equations the following expressions for $\mathcal{N}=\mathcal{N}(\lambda,\,a)$ and $\gamma=\gamma(\lambda,\,a)$ are respectively obtained:
\begin{align}
\mathcal{N}&=\pi\left[\frac{\gamma}{b}(a+b)^{2}+ 2\int_{A}^{B}\,w_{d}\,I_{0\lambda}R\,dR\,\right]\,\,\,\,\,\,\,\,\text{and}\,\,\,\,\,\,\,\,\gamma=-\frac{b}{\lambda(a+b)}\int_{A}^{B}\,w_{d}\,I_{0a}R\,dR, \label{gamforce}
\end{align}
where $w_{d}=w'(I_{0})$, $w_{dd}=w''(I_{0})$ etc., $I_{0\lambda}=\partial I_{0}/\partial\lambda$ and $I_{0a}=\partial I_{0}/\partial a$. We note that the $\gamma$ in $(\ref{gamforce})_{1}$ is eliminated through substitution of $(\ref{gamforce})_{2}$. For any loading scenario (e.g. fixed surface tension with monotonically varying axial stretch, or fixed axial stretch with increasing surface tension), the condition for localised bifurcation is \cite{EmeryFu2020,fu2016localized}
\begin{align}
\mathcal{J}(\lambda,\,a)\equiv\frac{\partial\mathcal{N}}{\partial a}\,\frac{\partial\gamma}{\partial \lambda} - \frac{\partial \mathcal{N}}{\partial \lambda}\,\frac{\partial \gamma}{\partial a}=0. \label{case1bifcon}
\end{align}
Now, it was originally thought in \cite{EmeryFu2020} that all values of $\lambda$ and $a$ which satisfy $(\ref{case1bifcon})$ are associated with negative values of $\gamma=\gamma(\lambda,a)$. Given that negative surface tension is physically implausible, it was concluded that localisation is not possible in this case. However, on re-examination of the contours $\mathcal{J}(\lambda,\,a)=0$ in the $(\lambda,a)$ plane, we find that there is in-fact an inconspicuous lower branch present in the regime of extremely small $a$ in addition to the main branch that we originally identified. It transpires that the values of $\lambda$ and $a$ along this branch correspond to \textit{positive} values of $\gamma$, and so contrary to our initial thoughts, localisation is theoretically possible in Case 1. However, it was shown in \cite{emery2021elasto} that the tube will bifurcate into an elliptic circumferential buckling mode at a far lower value of $\gamma$ than is predicted from $(\ref{gamforce})-(\ref{case1bifcon})$ for localisation, suggesting that the latter won't actually occur in reality. As an example, for the neo-Hookean strain-energy with $A=0.4$ and $\lambda=1.7$ fixed, the critical surface tension at which bifurcation into a circumferential elliptic mode occurs is $\gamma_{\tx{cr}}\approx 0.056$ \cite{emery2021elasto}, whereas for localisation the corresponding value is $\gamma_{\tx{cr}}\approx 6.65$. Thus, in the rest of this paper our attention will be focused on Cases 2 and 3.

\subsection{Case 2 - Radially fixed inner lateral boundary free of surface tension \label{sec3b}}

In Case $2$, the restrictions imposed on the inner lateral surface require that the deformed inner radius is unchanged from its initial value. That is, we must enforce the constraint $a=A$. The total energy of the primary state is then the same as $(\ref{TPEcase1})$, except we must set $\mathcal{L}_{s}^{A}=0$ since there is no surface tension on the inner lateral boundary here. Since $a$ is known in Case 2, the single parameter $\lambda$ is sufficient to determine the deformation completely, and equilibrium requires only that $\partial \mathcal{E}/\partial \lambda=0$. Say we fix the surface tension $\gamma$ and monotonically vary $\lambda$ from some initial value, then this equilibrium equation yields an expression $\mathcal{N}=\mathcal{N}(\lambda)$ which is simply $(\ref{gamforce})_{1}$ with $a\rightarrow A$ and $\gamma$ equal to the chosen fixed value.
 Alternatively, we may fix $\mathcal{N}$ and solve $\partial\mathcal{E}/\partial\lambda=0$ for $\gamma=\gamma\,(\lambda)$ instead. The condition for localised bifurcation then reduces from $(\ref{case1bifcon})$ to $d \mathcal{N}/d \lambda=0$ or $d \gamma /d \lambda = 0$, respectively. In the former case, the following implicit relationship between the critical axial stretch $\lambda_{\tx{cr}}$ for localisation and the fixed surface tension $\gamma$ is obtained:
\begin{align}
\gamma&=\frac{4\,b^{3}\lambda^{3}}{(A^{2}-1)^{2}}\int_{A}^{B}\frac{\partial}{\partial \lambda}\left(w_{d}\,I_{0\lambda}\right)R\,dR\,\bigg\vert_{\lambda=\lambda_{\tx{cr}}}. \label{bifconn}
\end{align}

\begin{figure}[h!]
\centering
\begin{subfigure}[t]{0.48\textwidth}
\includegraphics[width=\linewidth, valign=t]{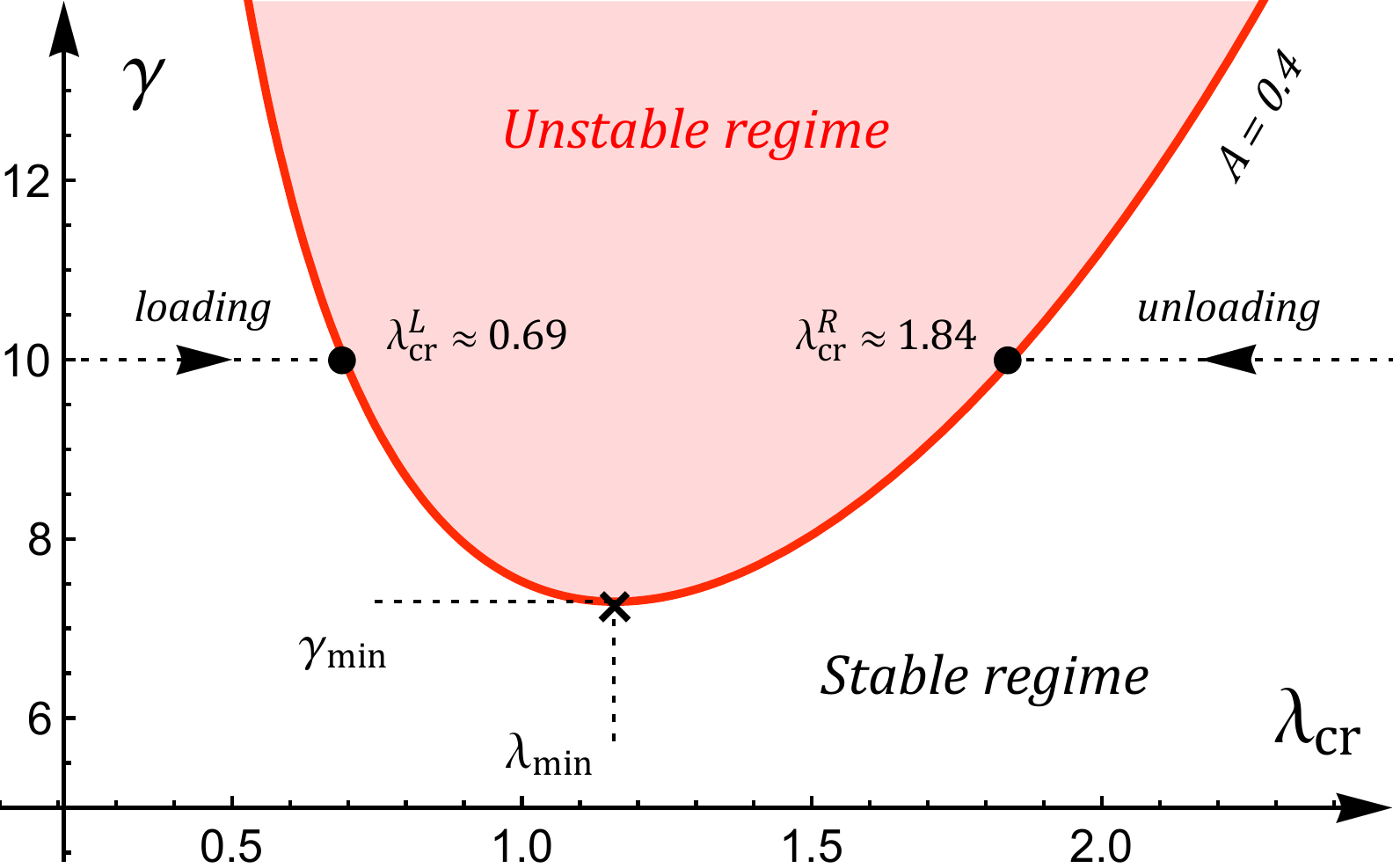}
\subcaption*{\textbf{(a)}}
\end{subfigure}\hfill
\begin{subfigure}[t]{0.4775\textwidth}
\includegraphics[width=\linewidth, valign=t]{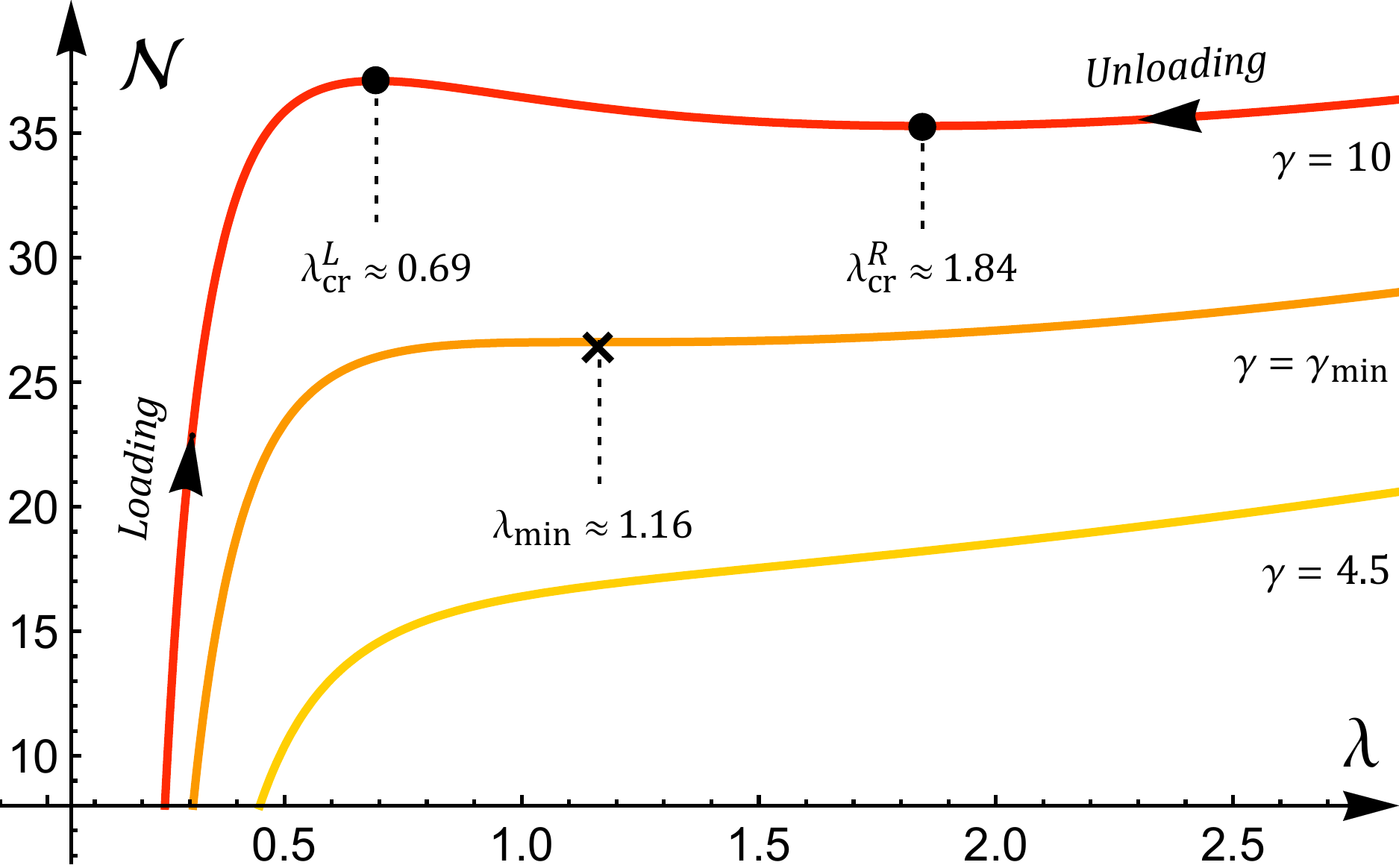}
\subcaption*{\textbf{(b)}}
\end{subfigure}
\caption{(Colour online) \textbf{(a)} The bifurcation condition $(\ref{bifconn})$ plotted in the $(\lambda_{\tx{cr}},\gamma)$ plane for the Gent material model $(\ref{neohook})_{2}$ with $J_{\tx{m}}=100$ and $A=0.4$. The bifurcation curve has a minimum at $\left(\lambda_{\tx{min}},\gamma_{\tx{min}}\right)\approx\left(1.16,7.3\right)$ as marked by the black cross. Then, for each fixed $\gamma>\gamma_{\tx{min}}$, there exists a bifurcation point either side of $\lambda=\lambda_{\tx{min}}$. For example, where $\gamma=10$, the tube can bifurcate into a localised solution at $\lambda_{\tx{cr}}^{L}\approx0.69<\lambda_{\tx{min}}$ and $\lambda_{\tx{cr}}^{R}\approx1.84>\lambda_{\tx{min}}$ as shown by the black dots. \textbf{(b)} The variation of $\mathcal{N}$ with respect to $\lambda$ for $\gamma=10$ (red), $\gamma=\gamma_{\text{min}}$ (orange) and $\gamma=4.5$ (yellow).}
\label{Fig2}
\end{figure}

An important feature of $(\ref{bifconn})$ is that the associated bifurcation curves in the $\left(\lambda_{\tx{cr}},\,\gamma\right)$ plane have a minimum at $(\lambda_{\text{min}},\,\gamma_{\text{min}})$, say. For each fixed $\gamma>\gamma_{\text{min}}$, there exists a single bifurcation value for $\lambda$ either side of $\lambda=\lambda_{\text{min}}$, say $\lambda_{\tx{cr}}^{L}<\lambda_{\text{min}}$ and $\lambda_{\tx{cr}}^{R}>\lambda_{\text{min}}$, which corresponds to the local maximum and minimum of $\mathcal{N}=\mathcal{N}(\lambda)$, respectively; see Fig. $\ref{Fig2}$. In the limit $\gamma\rightarrow\gamma_{\text{min}}$, these two extrema of $\mathcal{N}$ coalesce to form an inflection point, and for any $\gamma<\gamma_{\text{min}}$, $\mathcal{N}$ is a monotonically increasing function of $\lambda$  and localised bifurcation cannot occur.

Now, for any strain-energy function of the form $(\ref{SEfunction})$, $\mathcal{N}$ is a monotonic function of $\lambda$ up to the first bifurcation point encountered, which means that we can equivalently take $\mathcal{N}$ or $\lambda$ as the control parameter in this scenario. We may then consider two distinct approaches to varying $\mathcal{N}$ or $\lambda$, and these approaches are referred to as "loading" and "unloading" hereafter. When "loading", we fix $\gamma>\gamma_{\tx{min}}$ with zero axial force $\mathcal{N}$ initially, and an axial stretch $\lambda<\lambda_{\tx{cr}}^{L}$ is produced. As we increase $\lambda$ from this initial value, we transverse along the $\mathcal{N}=\mathcal{N}(\lambda)$ curve from left to right (as shown by the "loading" arrow in Fig. $\ref{Fig2}$ (b)). As such, we will always encounter the maximum of $\mathcal{N}$ at $\lambda=\lambda_{\text{cr}}^{L}$ first, and it is only the bifurcation solution corresponding to this maximum that is of interest in this scenario.  When "unloading", we apply a dead load to an end of the tube such that an initial axial stretch $\lambda>\lambda_{\tx{cr}}^{R}$ is produced. We can then theoretically decrease the axial load until the tube bifurcates into the solution corresponding to the minimum of $\mathcal{N}$ at $\lambda=\lambda_{\text{cr}}^{R}$. It is noted however that this approach lacks physical viability since it is somewhat unrealistic to expect soft slender tubes to withstand such a dead load.

\subsubsection{A spectral interpretation}
\begin{figure}[h!]
\centering
\includegraphics[scale=0.5]{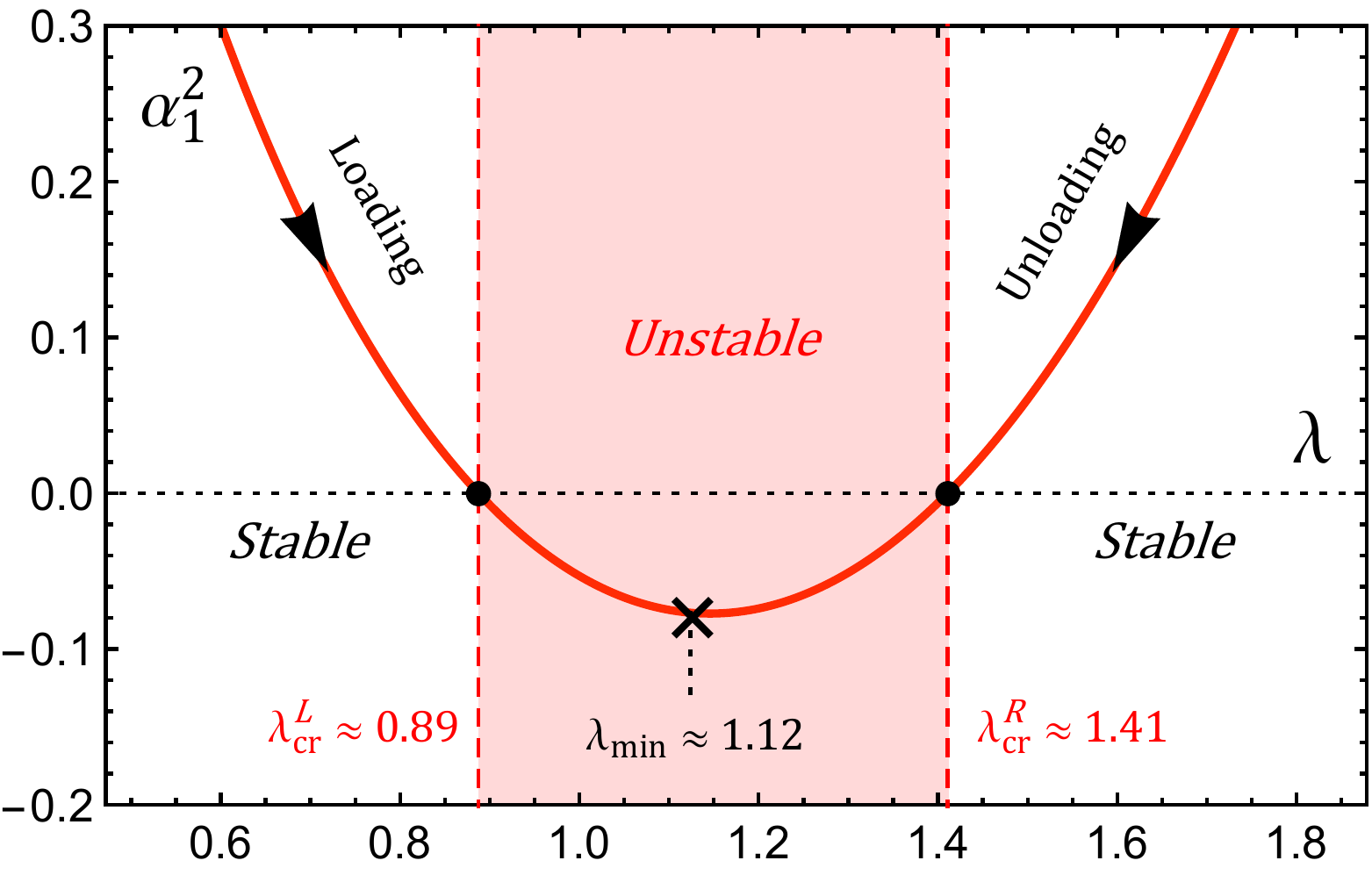}
\caption{The variation of $\alpha_{1}^{2}$ with respect to $\lambda$ for $A=0.5$ and $\gamma=10$.}
\label{Fig3}
\end{figure}

The analysis presented previously can be extended by a spectral approach. On taking $\lambda$ as the control parameter with $\gamma$ fixed, we enforce the ansatz $\phi=\phi_{0}+\varepsilon\,f(R)\,e^{\alpha z}$, where $\varepsilon\ll 1$ and $\alpha$ is the spectral parameter to be determined. We substitute this solution firstly into $(\ref{goveqn})$, $(\ref{BC1B})$ and $(\ref{BC2})$ and linearise in terms of $f$. We also require that the incremental radial displacement vanishes on $R=A$, i.e. $f(A)=0$; see $(\ref{incphi})_{1}$. We obtain the following linear eigenvalue problem:
\begin{align}
\frac{d\bm{f}}{d R}&=\textsf{A}\,(R,\alpha)\bm{f},\,\,\,\,\,\,\,\,\textsf{B}_{1}(A,\alpha)\bm{f}=\bm{0},\,\,\,\,\,\,\,\,\textsf{B}_{2}(B,\alpha)\bm{f}=\bm{0}, \label{eigsys}
\end{align}
where $\bm{f}=[f,\,f',\,f'',\,f''']^T$ and the matrices $\textsf{A},\,\textsf{B}_{1}$ and $\textsf{B}_{2}$ can be obtained from the supplementary \textit{Mathematica} code. The eigensystem $(\ref{eigsys})$ can be solved numerically through a determinant shooting method, and we refer the reader to \cite{EmeryFu2020} for details of its implementation. It is found that, on fixing $\gamma$, the system $(\ref{eigsys})$ has a trivial eigenvalue $\alpha=0$, infinitely many complex eigenvalues and infinitely many real eigenvalues $\alpha=\pm\alpha_{1},\,\pm\alpha_{2},\,\pm\alpha_{3}\dots$, where $|\alpha_{1}|<|\alpha_{2}|<|\alpha_{3}|<\cdots$. In Fig. $\ref{Fig3}$, we study the change in $\alpha_{1}^{2}$ for a representative case as we monotonically vary the load parameter $\lambda$.

Say that, on fixing $\gamma>\gamma_{\tx{min}}$, we take $\mathcal{N}=0$, producing an initial axial stretch $\lambda<\lambda_{\tx{cr}}^{L}$. As $\lambda$ is increased from this value, we transcend along the red curve in Fig. $\ref{Fig3}$ in the direction of the arrow marked "loading". We observe that $\alpha_{1}^{2}$ is positive, and hence $\pm\alpha_{1}$ are non zero and real, up to the bifurcation point $\lambda=\lambda_{\tx{cr}}^{L}$ (as indicated by the left-most black dot), where $\alpha_{1}^{2}=0$. Therefore, bifurcation into a localised solution coincides with zero becoming a triple eigenvalue of the system $(\ref{eigsys})$, and axial stretches below this bifurcation value lie in the sub-critical or stable regime wherein $\pm\alpha_{1}$ are real. Beyond this bifurcation value, we enter the unstable regime, and we see that $\alpha_{1}^{2}$ becomes negative, with $\pm\alpha_{1}$ therefore becoming purely imaginary. Thus, it is only \textit{after} localised bifurcation occurs that the eigensystem $(\ref{eigsys})$ can support periodic solutions, and this was also shown to be the case in \cite{EmeryFu2020} for the alternate loading scenario of fixed $\lambda>1$ and varying $\gamma$. Notwithstanding, if $\mathcal{N}$ is fixed such that $\lambda>\lambda_{\tx{cr}}^{R}$ initially and we unload from this point, the previous interpretations are also still valid. However, we transcend along the curve in Fig. $\ref{Fig3}$ in the direction of the "unloading" arrow, and localisation occurs at the right-most black dot instead.

\subsection{Case 3 - Radially fixed outer lateral boundary free of surface tension \label{sec3c}}

In Case 3, since the radial displacement of the \textit{outer} lateral surface is instead prohibited, the constraint $b=B$ is enforced, and we may then determine that the inner deformed radius $a=\sqrt{\lambda^{-1}(A^{2}-1)+1}$. To ensure that this expression for $a$ is real, we require that $\lambda> 1-A^{2}$, and in the limit $\lambda\rightarrow 1-A^{2}$, the inner deformed radius $a\rightarrow 0$. The total energy of the primary solution is again slightly modified from $(\ref{TPEcase1})$ since $\mathcal{L}_{s}^{B}$ must be zero. Then, for fixed $\gamma$, the resultant axial force $\mathcal{N}$ can be obtained from the equilibrium equation $\partial \mathcal{E}/\partial \lambda =0$, and is found to be equivalent to $(\ref{gamforce})_{1}$ but with the expressions for $a$ and $b$ in Case 3 substituted. The corresponding localised bifurcation condition is again $d \mathcal{N}/d \lambda=0$, from which we obtain
\begin{align}
\gamma&=\frac{4\,a^{3}\lambda^{3}}{(A^{2}-1)^{2}}\int_{A}^{B}\frac{\partial}{\partial \lambda}\left(w_{d}\,I_{0\lambda}\right)R\,dR\,\bigg\vert_{\lambda=\lambda_{\tx{cr}}}. \label{bifconn3}
\end{align}
If $\mathcal{N}$ is zero when applying the fixed surface tension, an axial stretch $1-A^{2}<\lambda<1$ is produced. For larger fixed $\gamma$, the inner radius of the tube will be smaller.

The weakly non-linear analysis in the following section is tailored towards Case 2. However, an overview of the results for Case 3 (and their distinctions from Case 2) is given in section $\ref{sec6}$ for completeness. In the next section, we not only validate $(\ref{bifconn})$, but we also demonstrate that the bifurcation solutions corresponding to the local maximum and minimum of $\mathcal{N}$ are sub-critical and are explicitly localised necking and bulging, respectively.
\section{Weakly non-linear near-critical analysis \label{sec4}}
In a weakly non-linear analysis, we are interested in the relationship between the increment of the control parameter from its bifurcation value and the amplitude of the associated localised solution. Further insights into the principal ideas of such an analysis can be found in \cite{yibin2001}. Here, we construct an exhaustive weakly non-linear analysis for Case 2 in terms of a general material model and focus on two main loading scenarios. Namely, we fix $\gamma$ and take $\lambda$ as the control parameter, or we fix $\lambda$ and take $\gamma$ as the control parameter.
\subsection{Taking $\lambda$ as the control parameter with fixed $\gamma$. \label{sec4a}}
Guided by the framework in \cite{yibin2001}, we consider a small deviation of the axial stretch from its critical value for localisation $\lambda_{\tx{cr}}$:
\begin{align}
\lambda&=\lambda_{\tx{cr}}+\varepsilon\,\lambda_{1}, \label{laminc}
\end{align}
where $\varepsilon\ll 1$ is a positive parameter and $\lambda_{1}$ is a constant of $O(1)$. From the spectral analysis in section $\ref{sec3}$ $\ref{sec3b}$ (i), we find that $\lambda$ is parabolic with respect to the axial wavenumber $k=-i\,\alpha_{1}$ in this near-critical regime, motivating the introduction of a far distance variable $s$ such that
\begin{align}
s&=\varepsilon^{1/2}\,z. \label{s}
\end{align}
Again, guided by \cite{yibin2001}, we extend $(\ref{phi0gen})$ and look for a solution of the form
\begin{align}
\phi&=\phi_{0} + \varepsilon^{1/2}\left\{\phi_1^{(1)}(R,s) + \varepsilon\,\phi_1^{(2)}(R,s) + \varepsilon^2\,\phi_1^{(3)}(R,s)+\dotsm\right\}. \label{phiexp}
\end{align}
Then, the corresponding expansions for the mixed co-ordinates given in $(\ref{incphi})$ are as follows:
\begin{align}
r&=r_{\tx{cr}} + \frac{\varepsilon}{r_{\tx{cr}}}\left\{\frac{\lambda_{1}}{2\lambda_{\tx{cr}}^{2}}\left(A^2 - R^2\right)+\phi_{1,s}^{(1)}\right\}+\dotsm,\,\,\,\,\,\,\,\,Z=\frac{z}{\lambda_{\tx{cr}}} + \frac{\varepsilon^{1/2}}{R}\phi_{1,R}^{(1)}-\varepsilon\frac{z\lambda_{1}}{\lambda_{\tx{cr}}^{2}}+\dotsm, \label{disp}
\end{align}
where $r_{\tx{cr}}$ is simply $r_{0}$ as given by $(\ref{rR})$ evaluated at $\lambda=\lambda_{\tx{cr}}$. On substituting $(\ref{phiexp})$ into $(\ref{goveqn})$ and the associated boundary conditions, we obtain a hierarchy of boundary value problems by equating the coefficients of like powers of $\varepsilon$. To leading order, we obtain the governing equation
\begin{align}
\mathcal{L}\left[\phi^{(1)}_1\right]&=0, \label{O1GE}
\end{align}
and the associated boundary conditions
\begin{align}
\mathcal{B}_{1}\left[\phi^{(1)}_1\right]_{R=1}&=0,\,\,\,\,\,\,\,\,
\mathcal{B}_{2}\left[\phi^{(1)}_1\right]_{R=1,\,A}=0,\,\,\,\,\,\,\,\,
\phi^{(1)}_{1,s}\left(A,s\right)=0, \label{O1BCs}
\end{align}
with the three differential operators $\mathcal{L}$, $\mathcal{B}_{1}$ and $\mathcal{B}_{2}$ being defined as
\begin{align}
\mathcal{L}&=\frac{\partial }{\partial R}\frac{1}{R}\frac{\partial}{\partial R}R\,w_{d}\frac{\partial }{\partial R}\frac{1}{R}\frac{\partial}{\partial R},\,\,\,\,\,\,\,\,
\mathcal{B}_{1}=\frac{1}{R}\frac{\partial}{\partial R}R\,w_{d}\frac{\partial }{\partial R}\frac{1}{R}\frac{\partial}{\partial R},\,\,\,\,\,\,\,\,
\mathcal{B}_{2}=\frac{\partial }{\partial R}\frac{1}{R}\frac{\partial}{\partial R}.
\end{align}
We note that $(\ref{O1BCs})_{1,2}$ are derived from $(\ref{BC1B})$ and $(\ref{BC2})$, respectively, whilst $(\ref{O1BCs})_{3}$ ensures that the leading order incremental radial displacement vanishes on $R=A$; see $(\ref{disp})_{1}$.
Through repeated integration of $(\ref{O1GE})$, the following general solution for $\phi_{1}^{(1)}$ is determined:
\begin{flalign}
&& \phi^{(1)}_1&=C_{1}(s) R^2+\,C_{2}(s)\,\xi_{2}(R)+C_{3}(s)\,\xi_{3}(R)+C_{4}(s),& \label{O1gensol} \\[1em]\text{where}
&&\,\,\,\,\,\,\,\,\,\,\xi_{2}(R)&=\int_{A}^{R}\,u\int_{A}^{u}\frac{t}{w_{d}}\,dt\,du\,\,\,\,\,\,\,\,\text{and}\,\,\,\,\,\,\,\,\xi_{3}(R)=\int_{A}^{R}\,u\int_{A}^{u}\frac{1}{t\,w_{d}}\,dt\,du.& \label{xi23}
\end{flalign}
In the above expressions, the variable $R$ in $w_{d}$ should be replaced by $t$ (i.e. $w_{d}=w'(I_{0}(t))$). On substituting $(\ref{O1gensol})-(\ref{xi23})$ into $(\ref{O1BCs})_2$, we find that $C_{2}$ and $C_{3}$ must necessarily be zero, whilst $(\ref{O1BCs})_1$ is automatically satisfied and $(\ref{O1BCs})_3$ requires that $C_{4}'(s)=-A^2\,C_{1} '(s)$. We may integrate the latter equation with respect to $s$ and set the additive constant to zero without loss of generality since the displacements $(\ref{disp})$ depend only on the partial derivatives of $\phi_{1}^{(1)}$. This statement also holds true at higher orders. Thus, the particular leading order solution is
\begin{align}
\phi^{(1)}_1&=C_{1}(s)\left(R^2 - A^2\right), \label{phi11}
\end{align}
where $C_{1}(s)$ is to be determined.

At the next order order, we have the following governing equation:
\begin{align}
\mathcal{L}\left[\phi^{(2)}_{1}\right]&=p_{1}(R)\,C_{1}''(s),\,\,\,\,\,\,\,\,\text{where}\,\,\,\,\,\,\,\,p_{1}=p_{1}^{(1)}\,w_{d}+p_{1}^{(2)}\,w_{dd}+p_{1}^{(3)}\,w_{ddd}, \label{OepGE}
\end{align}
with the functions $p_{1}^{(m)}(R)$ $(m=1,2\,\,\text{and}\,\,3)$ available in the supplementary \textit{Mathematica} code. The associated normal and shear traction-free boundary conditions take the respective forms:
\begin{equation}
\begin{aligned}
\mathcal{B}_{1}\left[\phi^{(2)}_1\right]_{R=1}&=k_{1}C_{1}''(s),\,\,\,\,\,\,\,\,\mathcal{B}_{2}\left[\phi^{(2)}_1\right]_{R=1,A}=s_{1}(R)\,C_{1}''(s)\,\big\vert_{R=1,A}, \label{OepBC}
\end{aligned}
\end{equation}
where the constant $k_{1}=(k_{1}^{(1)}+k_{1}^{(2)}w_{d}+k_{1}^{(3)}w_{dd})\vert_{R=1}$ is likewise given in our \textit{Mathematica} code and $s_{1}(R)=R(R^2-A^2)/(r_{\text{cr}}\lambda_{\text{cr}})^2$. Furthermore, the boundary condition enforcing zero incremental radial displacement at $R=A$ at this order is $\phi_{1,s}^{(2)}(A,s)=0$. A general solution to $(\ref{OepGE})$ is
\begin{align}
\phi^{(2)}_1&=D_{1}(s)\,R^2+D_{2}(s)\,\xi_{2}(s)+D_{3}(s)\,\xi_{3}(s)+D_{4}(s)+C_{1}''(s)\,\mathcal{P}(R), \label{Oepgensol}
\end{align}
where $\mathcal{P}(R)$ is a particular integral given by
\begin{align}
\mathcal{P}(R)&=\int_{A}^{R}\,x\int_{A}^{x}\,\frac{1}{v\,w_{d}}\int_{A}^{v}\,u\,\int_{A}^{u} p_{1}(t)\,dt\,du\,dv\,dx. \label{PI}
\end{align}
We note that $\mathcal{P}(R)$ is non-elementary, and we evaluate it numerically using the procedure detailed in section 4 of \cite{ye2020weakly}. On substituting $(\ref{Oepgensol})$ into $\phi_{1,s}^{(2)}(A,s)=0$ and $(\ref{OepBC})_{2}$, we find that $D_{2}$, $D_{3}$ and $D_{4}$ are linear in terms of $D_{1}$ and $C_{1}''$. Then, on substituting $(\ref{Oepgensol})$ into $(\ref{OepBC})_1$, the resulting equation can be shown to be numerically equivalent to the bifurcation condition $(\ref{bifconn})$ when the Gent strain-energy $(\ref{neohook})_{2}$ is deployed. In contrast, for the neo-Hookean model $(\ref{neohook})_1$, $(\ref{bifconn})$ can in-fact be recovered in closed form.

At the third order, the governing equation is
\begin{align}
\mathcal{L}\left[\phi^{(3)}_{1}\right]&=p_{1}(R)\,D_{1}''(s)+p_{2}(R)\,C_{1}''''(s) +p_{3}(R)\,C_{1}''(s)\left(\lambda_{1}-2\,\lambda_{\tx{cr}}^{2}\,C_{1}'(s)\right). \label{TOGE}
\end{align}
The normal and shear traction-free boundary conditions also contain non-linear inhomogeneous terms at this order, and they are expressed respectively as follows:
\begin{align}
\mathcal{B}_1\left[\phi_1^{(3)}\right]_{R=1}=\left\{k_{1}D_{1}''(s) + k_{2}\,C_{1}''''(s) + k_{3}\,C_{1}''(s)\left(\lambda_{1}-2\,\lambda_{\tx{cr}}^{2}\,C_{1}'(s)\right)\right\}\bigg\vert_{R=1},
\end{align}
and
\begin{align}
\mathcal{B}_{2}\left[\phi_{1}^{(3)}\right]_{R=1,A}=\left\{s_{1}\,D_{1}''(s)  +  s_{2}\,C_{1}''''(s) + s_{3}\,C_{1}''(s)\left(\lambda_{1}-2\,\lambda_{\tx{cr}}^{2}\,C_{1}'(s)\right)\right\}\bigg\vert_{R=1,A}.\label{TOBC}
\end{align}
Also, the condition enforcing zero incremental radial displacement on $R=A$ is $\phi_{1,s}^{(3)}(A,s)=0$. The expressions for $p_{2,3}$, $s_{2,3}$ and $k_{2,3}$ are lengthy, which makes obtaining the desired amplitude equation by solving the above third order boundary value problem algebraically and computationally involved. Alternatively, since the homogeneous form of this third order boundary value problem has a non-trivial solution, we may use the fact that the inhomogeneous terms on the right hand side of $(\ref{TOGE})$ - $(\ref{TOBC})$ must satisfy a solvability condition. In-fact, for sufficiently smooth functions $f(R)$ and $g(R)$, the following identity holds true:
\begin{align}
\int_{A}^{1}\left\{\,g\,\mathcal{L}\left[\,f\,\right]-f\,\mathcal{
L}\left[\,g\,\right]\,\right\}\,dR=\left[\,g\,\mathcal{B}_1\left[\,f\,\right]-f\,\mathcal{B}_1\left[\,g\,\right]+f'\,w_{d}\,\mathcal{B}_{2}\left[\,g\,\right]-g'\,w_{d}\,\mathcal{B}_{2}\left[\,f\,\right]\,\right]_{A}^{1}, \label{SAidentity}
\end{align}
and originates from the self-adjointness of $\mathcal{L}$ \cite{ye2020weakly}. The additional constraints $f(A)=g(A)=0$ must also be enforced to ensure that terms in $(\ref{SAidentity})$ involving $\mathcal{B}_{1}\,[\,\,\cdot\,\,]$ from the normal traction-free condition on $R=1$ are not evaluated at $R=A$. Particularly, on setting $g$ equal to the first order solution $(\ref{phi11})$ and $f=\phi_{1}^{(m)}$ $(m=2,\,3)$, $(\ref{SAidentity})$ reduces to
\begin{align}
\int_{A}^{1}\left(R^2 - A^2\right)\mathcal{L}\left[\,\phi_{1}^{(m)}\,\right]\,dR&=\left[\,\left(R^2-A^2\right)\mathcal{B}_{1}\left[\,\phi_{1}^{(m)}\,\right]-2\,R\,w_{d}\,\mathcal{B}_{2}\left[\,\phi_{1}^{(m)}\,\right]\,\right]_{A}^{1}, \label{RSAC}
\end{align}
and we note that $\mathcal{L}\,[\,\phi_{1}^{(m)}\,]$, $\mathcal{B}_{1}[\,\phi_{1}^{(m)}\,]$ and $\mathcal{B}_{2}[\,\phi_{1}^{(m)}\,]$ are each equal to expressions which involve only lower order solutions. On setting $m=2$ in $(\ref{RSAC})$ and equating coefficients of $C_{1}''$, we obtain the bifurcation condition for localisation. This condition is found to be numerically equivalent to $(\ref{bifconn})$ when the Gent material model is deployed, giving further verification of our derivations. On setting $m=3$ in $(\ref{RSAC})$, we yield the desired amplitude equation. Through integrating once and setting the arbitrary constant to zero for decay solutions, we obtain
\begin{align}
\scr{A}''&=\lambda_{1}\kappa_{1}\scr{A}+\kappa_{2}\,\scr{A}^{2}, \label{Ampeqn}
\end{align}
where the amplitude $\scr{A}(s)=C_{1}'(s)$ and the coefficients $\kappa_{1,2}$ are discussed below.
\subsubsection{Analysis of the amplitude equation}
For any strain-energy function of the form $(\ref{SEfunction})$, the special relationship $\kappa_{2}=-\lambda_{\tx{cr}}^{2}\,\kappa_{1}$ is found to hold and can be explained as follows. On substituting $(\ref{phiexp})$ into the $zZ$ component of $\vec{F}$, the following expansion of the principal axial stretch is determined to $O(\varepsilon)$:
\begin{align}
\lambda=\lambda_{\tx{cr}}+\varepsilon\left(\lambda_{1}-2\lambda_{\tx{cr}}^{2}\,\scr{A}(s)\right). \label{lamex2}
\end{align}
As is now established, the bifurcation points $\lambda_{\text{cr}}=\lambda_{\text{cr}}^{L}$ and $\lambda_{\text{cr}}=\lambda_{\tx{cr}}^{R}$ for localisation occur respectively at the local maximum and minimum of the resultant axial force $\mathcal{N}$ when $\gamma>\gamma_{\text{min}}$ is fixed. Therefore, $\mathcal{N}$ must admit parabolic behaviour in a small neighbourhood of $\lambda_{\tx{cr}}$ and, provided the amplitude $\scr{A}(s)$ is constant and non-zero, $(\ref{laminc})$ and $(\ref{lamex2})$ are two distinct near-critical solutions which must be equidistant from $\lambda_{\tx{cr}}$ and yield the same value of $\mathcal{N}$. That is, we must have
\begin{align}
\lambda_{\tx{cr}}-\left\{\lambda_{\tx{cr}}+\varepsilon\left(\lambda_{1}-2\lambda_{\tx{cr}}^{2}\,\scr{A}(s)\right)\right\}&=\left(\lambda_{\tx{cr}}+\varepsilon\lambda_{1}\right)-\lambda_{\tx{cr}}, \label{ident}
\end{align}
from which we obtain $\scr{A}(s)=\lambda_{1}/\lambda_{\tx{cr}}^{2}$. Then, on substituting this expression for $\scr{A}$ back into $(\ref{Ampeqn})$, the relation $\kappa_{2}=-\lambda_{\tx{cr}}^{2}\,\kappa_{1}$ follows.

Whilst the determined expression for $\kappa_{1}$ is analytical, it is largely in terms of integrals which cannot be evaluated explicitly. Thus, for the chosen material model, $\kappa_{1}$ must be determined numerically by evaluating these integrals through the approach given in \cite{ye2020weakly}. Nevertheless, by the following interpretation, we expect that $\kappa_{1}$ is negative (resp. positive) for any $\lambda_{\text{cr}}=\lambda_{\text{cr}}^{L}<\lambda_{\text{min}}$ (resp. $\lambda_{\text{cr}}=\lambda_{\text{cr}}^{R}>\lambda_{\text{min}}$) such that $\gamma>\gamma_{\text{min}}$ is fixed. Consider the linearised form of the amplitude equation $(\ref{Ampeqn})$. On assuming a solution of the form $\scr{A}=e^{\alpha s}$, the spectral parameter $\alpha$ is found to take the non-trivial values $\pm\alpha_{1}=\pm\sqrt{\lambda_{1}\kappa_{1}}$. We note that these are the same $\pm\alpha_{1}$ from the section $\ref{sec3}$ $\ref{sec3b}$ (i) defined analytically in the near-critical regime. From the spectral analysis, we expect that bounded periodic solutions (i.e. purely imaginary values of $\pm\alpha_{1}$) are possible only in the regime unstable to localisation, and that $\pm\alpha_{1}$ are real in the stable regime. With reference to Fig. $\ref{Fig2}$, $\lambda_{1}$ is negative (resp. positive) in the stable (resp. unstable) regime when "loading" and we therefore require that $\kappa_{1}<0$ for $\pm\alpha_{1}$ to take its expected form. In contrast, $\lambda_{1}$ is positive (resp. negative) in the stable (resp. unstable) regime when "unloading", and so we must have $\kappa_{1}>0$ instead. On specifying the Gent strain-energy function with $J_{\tx{m}}=100$, we present in Tables $\ref{table1}$ and $\ref{table2}$ numerical values of $\kappa_{1}$ corresponding to $\lambda_{\tx{cr}}=\lambda_{\tx{cr}}^{L}$ ("loading") and $\lambda_{\tx{cr}}=\lambda_{\tx{cr}}^{R}$ ("unloading"), respectively. We observe that, in the former and latter cases, $\kappa_{1}$ is respectively negative and positive, suggesting that $\kappa_{1}$ does indeed change sign as we pass through $\lambda=\lambda_{\text{min}}$.
\begin{table}[h!]
\centering
\begin{threeparttable}
\caption{Numerical values of $\kappa_{1}$ for the Gent material model with $J_{\tx{m}}=100$ and $\lambda_{\tx{cr}}=\lambda_{\tx{cr}}^{L}<\lambda_{\text{min}}$ (i.e. when "loading").}
\begin{tabular}{c c c c c c c} 
\hline 
& $\gamma=10.5$ & $\gamma=11$ & $\gamma=11.5$ & $\gamma=12$ & $\gamma=12.5$ & $\gamma=13$ \\ [0.5ex] 
\hline 
$A=0.2$ & -1.2349 & -1.2548 & -1.2703 & -1.2823 & -1.2913 & -1.2977  \\
$A=0.3$ & -1.2226 & -1.2523 & -1.2766 & -1.2965 & -1.3128 & -1.3261  \\
$A=0.4$ & -1.1642 & -1.2108 & -1.2497 & -1.2825 & -1.3103 & -1.3341  \\
$A=0.5$ & -0.9856 & -1.0667 & -1.1336 & -1.1897 & -1.2375 & -1.2788 \\
$A=0.6$ & -0.2857 & -0.5763 & -0.7545 & -0.8698 & -0.9685 & -1.05002 \\ [1ex] 
\hline 
\end{tabular} \label{table1}
\end{threeparttable}
\end{table}
\begin{table}[h!]
\centering
\begin{threeparttable}
\caption{Numerical values of $\kappa_{1}$ for the Gent material model with $J_{\tx{m}}=100$ and $\lambda_{\tx{cr}}=\lambda_{\tx{cr}}^{R}>\lambda_{\text{min}}$ (i.e. when "unloading").}
\begin{tabular}{c c c c c c c} 
\hline 
& $\gamma=10.5$ & $\gamma=11$ & $\gamma=11.5$ & $\gamma=12$ & $\gamma=12.5$ & $\gamma=13$ \\ [0.5ex] 
\hline 
$A=0.2$ & 1.3889 & 1.4339 & 1.4748 & 1.5124 & 1.5472 & 1.5796  \\
$A=0.3$ & 1.3158 & 1.3632 & 1.4045 & 1.4433 & 1.4777 & 1.5092  \\
$A=0.4$ & 1.1925 & 1.2483 & 1.2965 & 1.3387 & 1.37604 & 1.4094  \\
$A=0.5$ & 0.96797 & 1.0495 & 1.1167 & 1.1735 & 1.2221 & 1.2643 \\
$A=0.6$ & 0.2717 & 0.5545 & 0.7158 & 0.8326 & 0.9237 & 0.9979 \\ [1ex] 
\hline 
\end{tabular} \label{table2}
\end{threeparttable}
\end{table}
A localised solution to the amplitude equation $(\ref{Ampeqn})$ is a standing solitary wave given by
\begin{align}
\scr{A}(s)&=-\frac{3\,\lambda_{1}\kappa_{1}}{2\,\kappa_{2}}\,\text{sech}^2\left(\frac{1}{2}\sqrt{\lambda_{1}\kappa_{1}}s\right). \label{As}
\end{align}

Given the aforementioned sign change of $\kappa_{1}$ across $\lambda=\lambda_{\text{min}}$, the form of $(\ref{As})$ means that this localised solution exists for $\lambda_{1}<0$ (resp. $\lambda_{1}>0$) when "loading" (resp. "unloading"). In other words, the solution $(\ref{As})$ emerges \textit{sub-critically} in both scenarios. Explicitly, $(\ref{As})$ is a dark soliton (necking) when "loading" and a bright soliton (bulging) when "unloading".


When the Gent material model is employed, we have verified numerically that $\kappa_{1}\rightarrow 0$ as $(\lambda_{\text{cr}},\gamma)\rightarrow(\lambda_{\tx{min}},\gamma_{\text{min}})$. Moreover, when reducing to the neo-Hookean model, it can be shown explicitly that $\kappa_{1}\propto d\gamma/d\lambda\vert_{\lambda_{\text{cr}}}$, where $\gamma$ is given by $(\ref{bifconn})$. Thus, the form of $(\ref{As})$ suggests that a rescaling of the dependent variable $s$ is required in this limit, and we analyse this case separately in the next subsection.

\subsubsection{The limit $\lambda_{\tx{cr}}\rightarrow\lambda_{\tx{min}}$}
By expanding $\kappa_{1}$ about $\lambda=\lambda_{\text{min}}$, the following re-scaling of $s$ in this limit is deduced from $(\ref{As})$:
\begin{align}
\hat{s}=\varepsilon^{1/2}s=\varepsilon\,z. \label{sh}
\end{align}
Since $\gamma$ is locally parabolic with respect to $\lambda_{\tx{cr}}$ near $\left(\lambda_{\tx{min}},\,\gamma_{\tx{min}}\right)$, we set
\begin{align}
\lambda&=\lambda_{\tx{min}}+\varepsilon\,\hat{\lambda}_{1}\,\,\,\,\,\,\,\,\text{and}\,\,\,\,\,\,\,\,\gamma=\gamma_{\tx{min}}+\varepsilon^{2}\,\hat{\gamma}_{1}, \la{lim1}
\end{align}
where $\hat{\lambda}_{1}$ and $\hat{\gamma}_{1}$ are constants of $O(1)$. Given $(\ref{sh})$, we implement the following re-scaling of $\phi$:
\begin{align}
\phi&=\phi_{0} + \hat{C}_{1}(\hat{s})\left(R^2 - A^2\right)+\varepsilon\,\hat{\phi}_{1}^{(1)} + \varepsilon^{2}\,\hat{\phi}_{1}^{(2)} + \varepsilon^{3}\,\hat{\phi}_{1}^{(3)} + \varepsilon^{4}\,\hat{\phi}_{1}^{(4)}+O(\varepsilon^{5}), \label{philim}
\end{align}
noting that, although the second term $\hat{C}_{1}(\hat{s})(R^2 - A^2)$ in  $(\ref{philim})$ is of the same order as the first term $\phi_{0}$, the corresponding deformation gradient is of higher order.

We then follow the same procedure as presented previously for the non-limit case. At $O(\varepsilon)$, we find that $\hat{\phi}_{1}^{(1)}$ must necessarily be zero. At $O(\varepsilon^{2})$, the bifurcation condition for localisation evaluated at $(\lambda_{\text{cr}},\gamma)=(\lambda_{\tx{min}},\gamma_{\tx{min}})$ is obtained and is found to be equivalent to $(\ref{bifconn})$ (numerically for the Gent model, but in closed form for the neo-Hookean model). Then, at $O(\varepsilon^{3})$, we recover an equation which is numerically equivalent to $d\gamma/d\lambda\,\vert_{\lambda_{\tx{min}}}=0$ (and which is satisfied automatically). At $O(\varepsilon^{4})$, by setting $g=\hat{C}_{1}(\hat{s})(R^{2}-A^{2})$ and $f=\hat{\phi}_{1}^{(4)}$ in $(\ref{RSAC})$, the following amplitude equation for $\hat{\scr{A}}(\hat{s})=\hat{C}_{1}'(\hat{s})$ is obtained:
\begin{align}
\hat{\scr{A}}''&=\left(\hat{\gamma}_{1}\hat{\kappa}_{1}+\hat{\lambda}_{1}^{2}\,\hat{\kappa}_{2}\right)\hat{\scr{A}}
-2\,\hat{\lambda}_{1}\lambda_{\tx{min}}^2\,\hat{\kappa}_{2}\,\hat{\scr{A}}^2+\frac{4}{3}\lambda_{\tx{min}}^4\hat{\kappa}_{2}\,\hat{\scr{A}}^3, \label{AEl}
\end{align}
where $\hat{\kappa}_{1}$ and $\hat{\kappa}_{2}$ are new constants.

Although we have initially derived two separate expressions for $\hat{\kappa}_{1}$ and $\hat{\kappa}_{2}$, simpler connections between these two coefficients can be established as follows. First, on substituting $\hat{\scr{A}}=e^{\hat{\alpha}_1\hat{s}}$ into the linearised form of $(\ref{AEl})$, we obtain $\hat{\alpha}_1^2= \hat{\gamma}_{1}\hat{\kappa}_{1}+\hat{\lambda}_{1}^{2}\,\hat{\kappa}_{2}$, which will be used repeatedly below. For fixed $\hat{\gamma}_{1}>0$, the local maximum and minimum of $\mathcal{N}$ are near the point of coalescence; see Fig. $\ref{Fig2}$. On Taylor expanding $\mathcal{N}$ around $(\lambda_{\tx{min}},\gamma_{\tx{min}})$, the following bifurcation points $\lambda_{\text{cr}}^{L,R}$ for localisation near $\lambda_{\tx{min}}$ can be deduced from the equation $d\mathcal{N}/d\lambda =0$:
\begin{align}
\lambda_{\tx{cr}}^{L,R}=\lambda_{\tx{min}}+\varepsilon\,\hat{\lambda}_{1}^{L,R},\,\,\,\,\,\,\,\,\text{where}\,\,\,\,\,\,\,\,\hat{\lambda}_{1}^{L,R}&=\mp\sqrt{-2\,\hat{\gamma}_{1}\frac{\partial^{2}\mathcal{N}}{\partial\lambda\partial\gamma}\left(\frac{\partial^{3}\mathcal{N}}{\partial \lambda^{3}}\right)^{-1}}, \label{lam1LR}
\end{align}
with the derivatives of $\mathcal{N}$ evaluated at $(\lambda_{\tx{min}},\gamma_{\tx{min}})$. When "loading", say, the bifurcation point of interest is $\lambda_{\text{cr}}^{L}=\lambda_{\tx{min}}+\varepsilon\,\hat{\lambda}_{1}^{L}$.  Since $\hat{\lambda}_{1}=\hat{\lambda}_{1}^{L}$ lie on the bifurcation curve, we must have $\hat{\alpha}_1=0$ when $\hat{\lambda}_{1}$ is replaced by $\hat{\lambda}_{1}^{L}$ in the above expression for $\hat{\alpha}_1^2$, which yields the connection
\begin{align}
\hat{\kappa}_{2}=-\frac{\hat{\gamma}_{1}\,\hat{\kappa}_{1}}{(\hat{\lambda}_{1}^{L})^2}. \label{k2hatt}
\end{align}
Alternatively, consider the following two term expansion of the original bifurcation condition $(\ref{bifconn})$ around $\gamma_{\text{min}}$:
\be
\gamma=\gamma_{\tx{min}}+\frac{1}{2}\,(\lambda-\lambda_{\tx{min}})^2 \cdot  \frac{d^2 \gamma}{d \lambda^2}\,\bigg\vert_{\lambda=\lambda_{\tx{min}}}. \la{adf1} \en
On substituting \rr{lim1} into \rr{adf1}, we obtain an expression for $\hat{\gamma}_{1}$. Since this also lies on the bifurcation curve, the corresponding $\hat{\alpha}_1$ must vanish as well, which then yields a second connection between $\hat{\kappa}_{1}$ and $\hat{\kappa}_{2}$:
\be
\hat{\kappa}_{2}=-\frac{1}{2} \hat{\kappa}_{1}  \frac{d^2 \gamma}{d \lambda^2}\, \bigg\vert_{\lambda=\lambda_{\tx{min}}}. \la{adf2} \en
As a consistency check, we have verified numerically that the two expressions \rr{k2hatt} and \rr{adf2} are indeed equivalent. These two expressions can further be used to show that $\hat{\kappa}_{1}$ must be negative whereas $\hat{\kappa}_{2}$ must be positive. To this end, we first substitute \rr{k2hatt} into the expression for $\hat{\alpha}_1^2$ to obtain $\hat{\alpha}_1=\pm\sqrt{\hat{\gamma}_{1}\hat{\kappa}_{1}(1-\hat{\lambda}_{1}^{2}/(\hat{\lambda}_{1}^{L})^{2})}$. 
In the stable regime where $\hat{\lambda}_{1}<\hat{\lambda}_{1}^{L}<0$, $\hat{\alpha}_1$ must be real as expected from the spectral analysis. It then follows that $\hat{\kappa}_{1}$ must necessarily be negative. In the unstable regime where $\hat{\lambda}_{1}^{L}<\hat{\lambda}_{1}<0$,  $\hat{\kappa}_{1}$ must remain negative for $\hat{\alpha}_1$ to be purely imaginary.  An analogous interpretation exists when "unloading", and the requirement that $\hat{\kappa}_{1}$ be negative remains true. The second connection \rr{adf2} then implies that $\hat{\kappa}_{2}$ is positive.

\begin{figure}[h!]
\centering
\begin{subfigure}[t]{0.483\textwidth}
\includegraphics[width=\linewidth, valign=t]{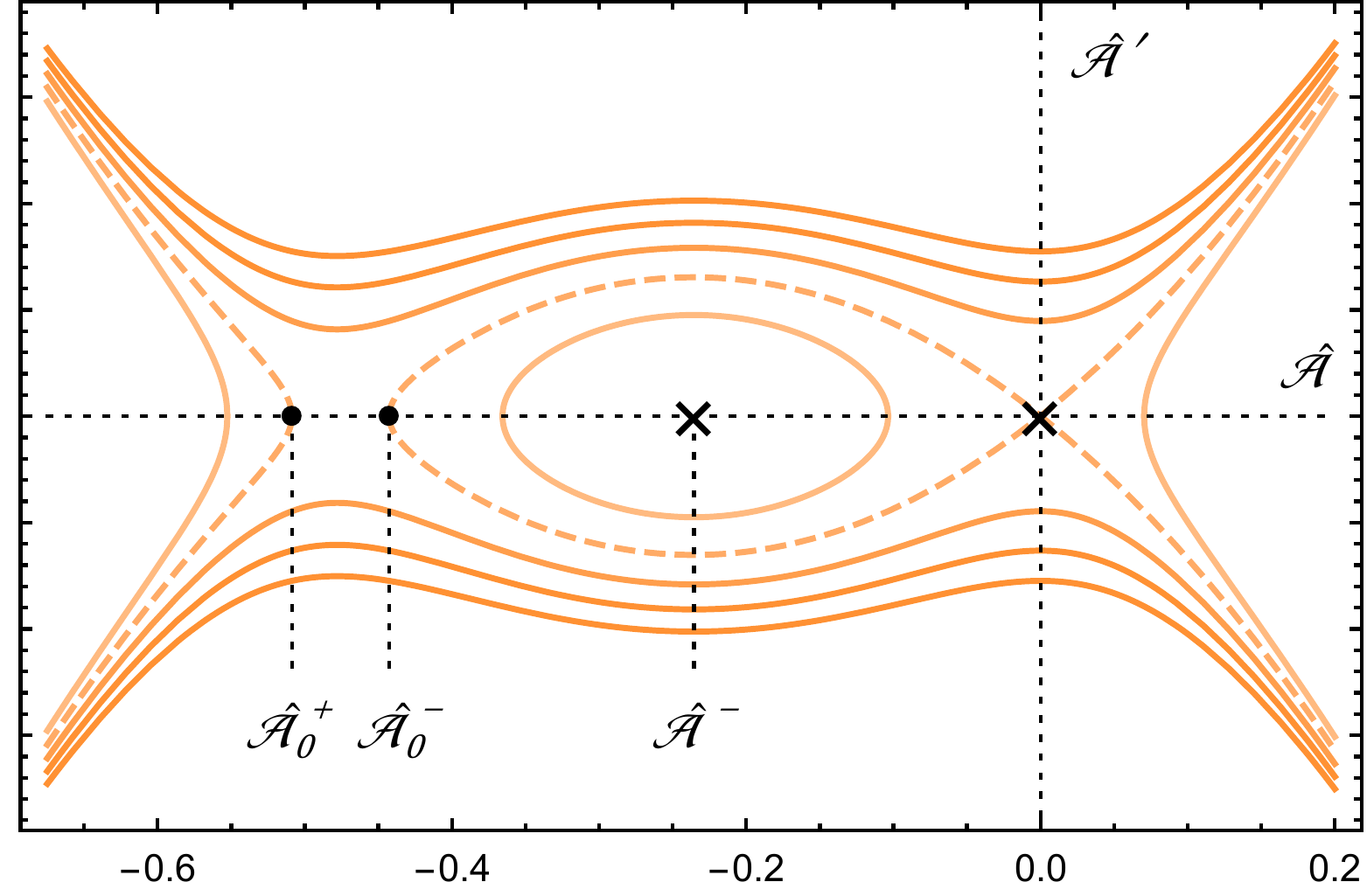}
\subcaption*{\textbf{(a)}}
\end{subfigure}\hfill
\begin{subfigure}[t]{0.483\textwidth}
\includegraphics[width=\linewidth, valign=t]{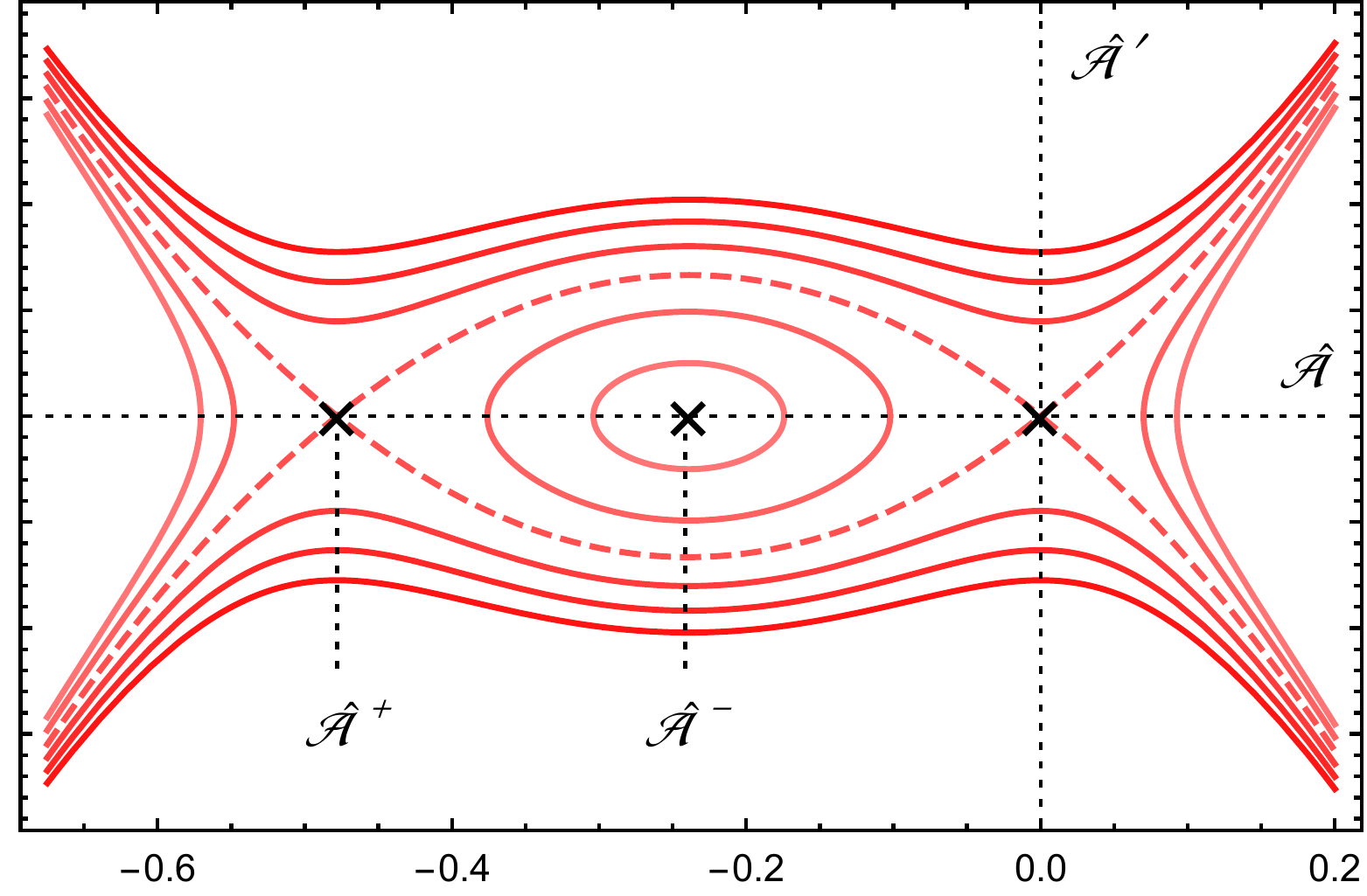}
\subcaption*{\textbf{(b)}}
\end{subfigure}
\caption{Phase portraits when $\hat{\lambda}_{1}$ is \textbf{(a)} close to and \textbf{(b)} equal to $\sqrt{3}\hat{\lambda}_{1}^{L}$. Away from $\hat{\lambda}_{1}=\sqrt{3}\hat{\lambda}_{1}^{L}$, a center $\hat{\scr{A}}^{-}$ exists inside a homoclinic orbit, with the latter connecting $\hat{\scr{A}}^{-}_{0}$ to the origin. In the limit $\hat{\lambda}_{1}\rightarrow\sqrt{3}\hat{\lambda}_{1}^{L}$, the ground states $\hat{\scr{A}}_{0}^{\pm}$ coalesce to form the saddle point $\hat{\scr{A}}^{+}$. This signals the transition from a localised (homoclinic) solution to a kink-wave (heteroclinic) solution.}
\label{Fig4}
\end{figure}

Alternatively, $(\ref{AEl})$ can be expressed as the one degree-of-freedom Hamiltonian system
\begin{align}
\hat{\scr{A}}''&=-\frac{\partial \hat{V}}{\partial \hat{\scr{A}}},\,\,\,\,\,\,\,\,\text{where}\,\,\,\,\,\,\,\,\hat{V}=-\frac{1}{3}\lambda_{\tx{min}}^{4}\,\hat{\kappa}_{2}\,\hat{\scr{A}}^{2}\left(\hat{\scr{A}}-\hat{\scr{A}}^{+}_{0}\right)\left(\hat{\scr{A}}-\hat{\scr{A}}^{-}_{0}\right). \label{AElim}
\end{align}
The fixed points $\hat{\scr{A}}^{\pm}$ and ground states $\hat{\scr{A}}_{0}^{\pm}$ of $(\ref{AElim})$ are then
\begin{flalign}
&&\hat{\scr{A}}^{\pm}&=\frac{3}{4}\left\{\,\frac{\hat{\lambda}_{1}}{\lambda_{\tx{min}}^2}\pm\frac{1}{\lambda_{\tx{min}}^2}\sqrt{\hat{\lambda}_{1}^2 - \frac{4}{3}\left(\hat{\lambda}_{1}^2 +\frac{\hat{\kappa}_{1}}{\hat{\kappa}_{2}}\hat{\gamma}_{1}\right)}\,\right\}& \label{FPs} \\[1em]\text{and}
&&\hat{\scr{A}}_{0}^{\pm}&=\frac{1}{\lambda_{\tx{min}}^2}\left\{\,\hat{\lambda}_{1}\,\pm\,\sqrt{-\frac{1}{2\hat{\kappa}_{2}}\left(3\hat{\kappa}_{1}\hat{\gamma}_{1}+\hat{\kappa}_{2}\hat{\lambda}_{1}^{2}\right)}\,\right\}.& \label{GSs}
\end{flalign}
Equation $(\ref{AElim})$ admits a localised solution if and only if the following two conditions are satisfied:
\begin{align}
\hat{\kappa}_{1}\hat{\gamma}_{1}+\hat{\kappa}_{2}\,\hat{\lambda}_{1}^{2}>0,\,\,\,\,\,\,\,\,3\,\hat{\kappa}_{1}\hat{\gamma}_{1}+\hat{\kappa}_{2}\,\hat{\lambda}_{1}^{2}\leq 0. \label{NCond}
\end{align}
The first condition $(\ref{NCond})_{1}$ ensures that the localised solution decays as $|\hat{s}|\rightarrow\infty$ and that there are fixed points other than $\hat{\scr{A}}=0$, whereas $(\ref{NCond})_{2}$ guarantees that $\hat{\scr{A}}'=0$ has a non-trivial root (i.e. $\hat{\scr{A}}_{0}^-$ is real). We observe that $(\ref{NCond})$ cannot be satisfied at $\hat{\lambda}_{1}=0$, and it will be shown shortly that a kink-wave solution exists in place of localisation at this point. On combining the inequalities in $(\ref{NCond})$, we obtain the following range of values of $\hat{\lambda}_{1}$ for which a localised solution can exist:
\begin{align}
\sqrt{3}\,\hat{\lambda}_{1}^{L}<\,\hat{\lambda}_{1}\,<\hat{\lambda}_{1}^{L}\,\,\,\,\,\,\,\,\text{and}\,\,\,\,\,\,\,\,\hat{\lambda}_{1}^{R}<\,\hat{\lambda}_{1}\,<\sqrt{3}\,\hat{\lambda}_{1}^{R}. \label{lam1int}
\end{align}
The localised solution is given explicitly in \cite{FuST} as follows:
\begin{align}
\hat{\scr{A}}(\hat{s})&=\frac{\hat{\scr{A}}_{0}^{+}\hat{\scr{A}}_{0}^{-}\,(1-\zeta^2)}{\hat{\scr{A}}_{0}^{+}-\hat{\scr{A}}_{0}^{-}\,\zeta^2},\,\,\,\,\,\,\,\,\text{where}\,\,\,\,\,\,\,\,\zeta(\hat{s})=\tanh\left[-\sqrt{\frac{\hat{\kappa}_{2}\hat{\scr{A}}_{0}^{+}\hat{\scr{A}}_{0}^{-}}{6}}\lambda_{\tx{min}}^{2}\hat{s}\right]. \label{ampsollim}
\end{align}
When "loading" (resp. "unloading"), this solution in the limit $\hat{\lambda}_{1}\rightarrow\hat{\lambda}_{1}^{L}$ (resp. $\hat{\lambda}_{1}\rightarrow\hat{\lambda}_{1}^{R}$) takes the form of a localised neck (resp. bulge), and a non-trivial fixed point $\hat{\scr{A}}^{-}$ (a center) emerges in this limit; see Fig. $\ref{Fig4}$ (a). However, in the limits $\hat{\lambda}_{1}\rightarrow\sqrt{3}\,\hat{\lambda}_{1}^{L}$ and $\hat{\lambda}_{1}\rightarrow\sqrt{3}\,\hat{\lambda}_{1}^{R}$, $(\ref{ampsollim})$ degenerates into a kink-wave solution characterised by two regions of uniform but distinct axial stretch
\begin{align}
\lambda_{L,R}&=\lambda_{\text{min}}+\sqrt{3}\,\varepsilon\,\hat{\lambda}_{1}^{L,R},
\end{align}
 connected by a smooth transition region. In this limit, the ground states $\hat{\scr{A}}_{0}^{\pm}$ coalesce to form the fixed point $\hat{\scr{A}}^{+}$ (a saddle) which connects to the origin via a heteroclinic orbit; see Fig. $\ref{Fig4}$ (b).

\begin{figure}[h!]
\centering
\begin{subfigure}[t]{0.495\textwidth}
\includegraphics[width=\linewidth, valign=t]{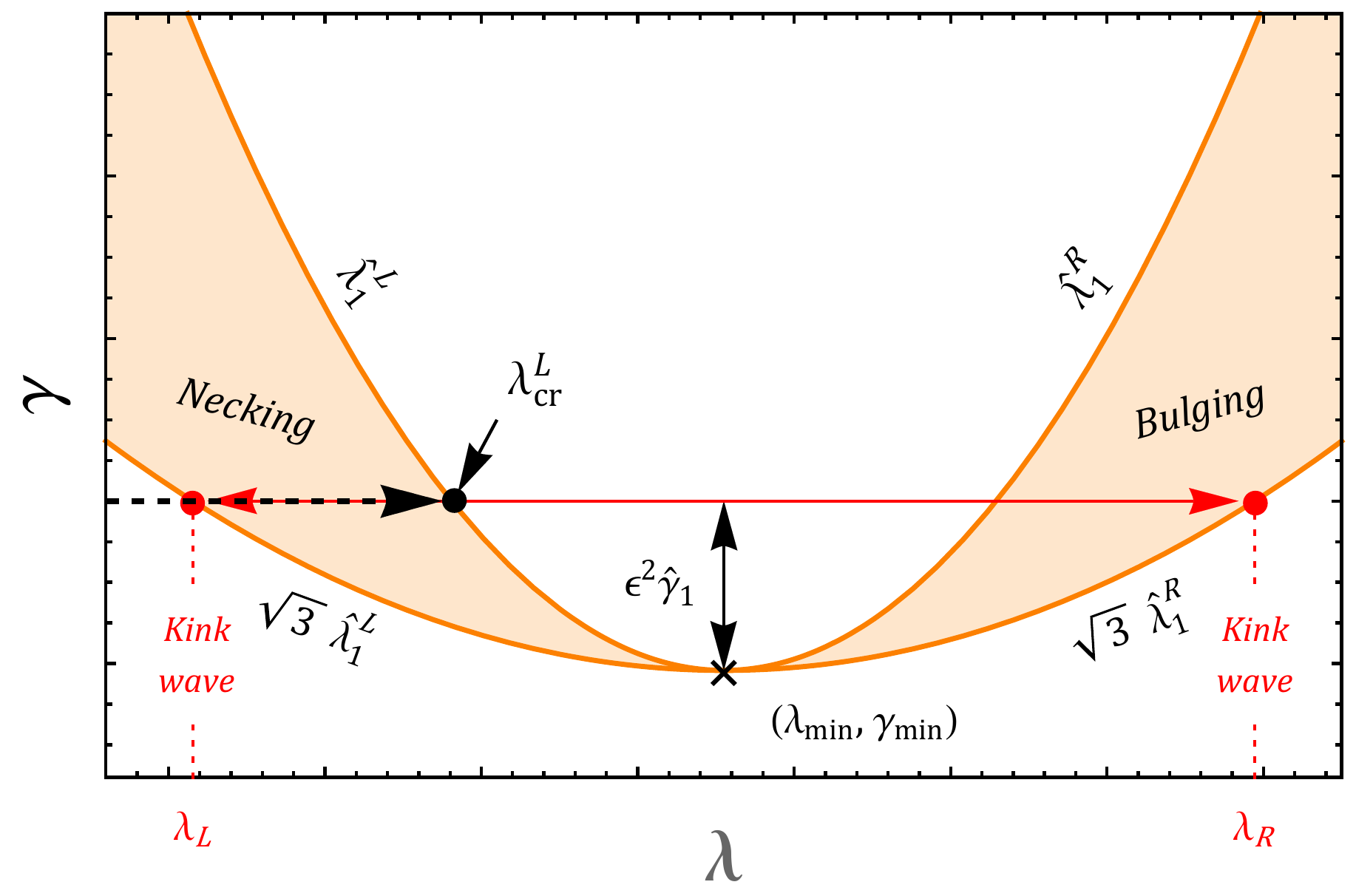}
\vspace{0.1mm}
\subcaption*{\textbf{(a)}}
\end{subfigure}\hfill
\begin{subfigure}[t]{0.475\textwidth}
\includegraphics[width=\linewidth, valign=t]{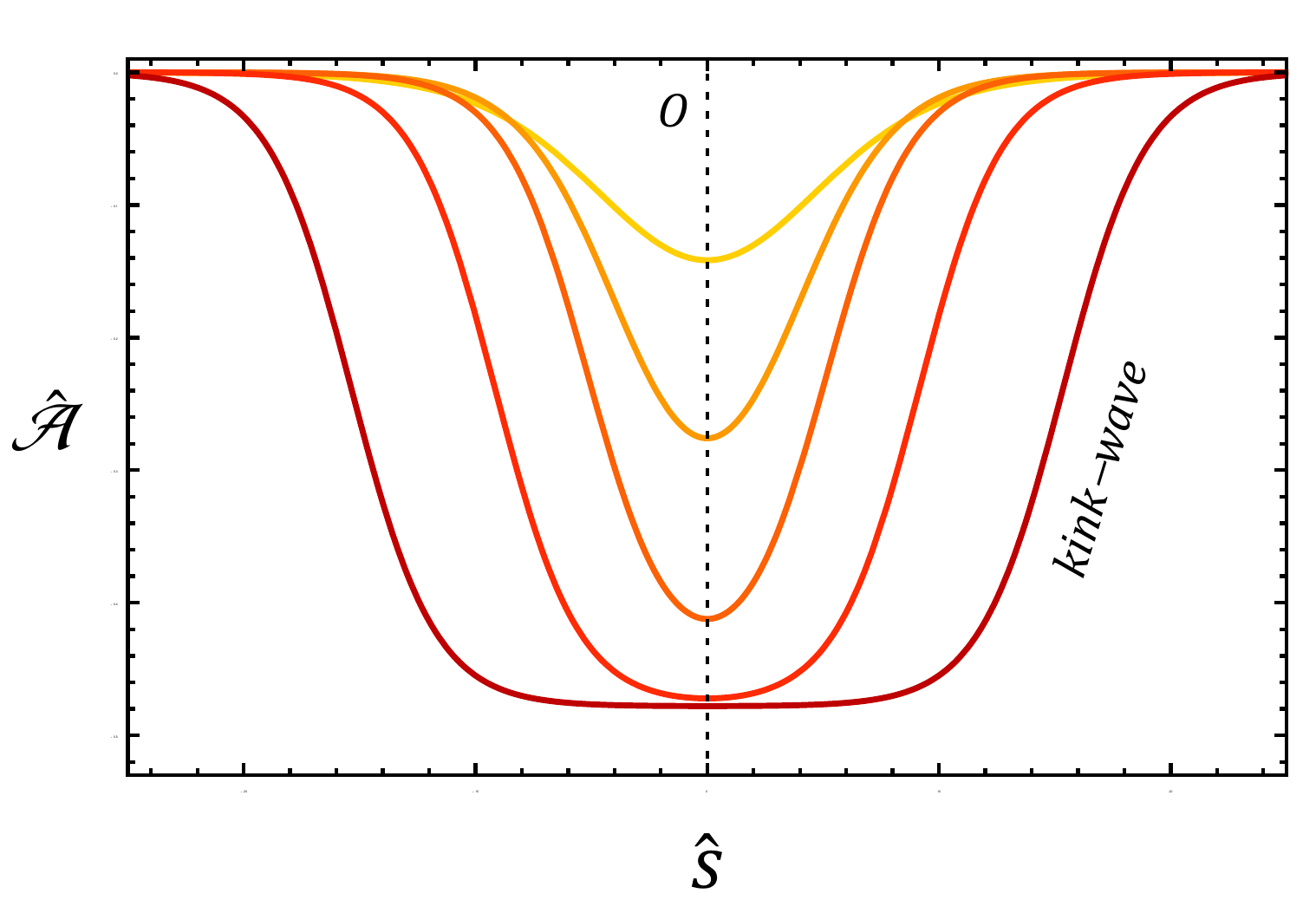}
\subcaption*{\textbf{(b)}}
\end{subfigure}
\caption{\textbf{(a)} A plot of the intervals of existence $(\ref{lam1int})$ of the localised solution $(\ref{ampsollim})$ (orange region). Say we fix $\hat{\gamma}_{1}>0$ with $\mathcal{N}=0$ initially (i.e. we enforce the "loading" scenario). Then, as $\lambda$ is increased, we move in the direction of the dashed black arrow shown. The initial bifurcation occurs sub-critically as we approach the black dot at $\lambda_{\text{cr}}^{L}$, and takes the form of a localised neck (as shown by the yellow curve in \textbf{(b)}). On increasing the overall stretch beyond $\lambda_{\text{cr}}^{L}$, a transition to a kink-wave solution as shown in \textbf{(b)} is expected, and the stretches $\lambda_{L,R}$ corresponding to the final "two-phase" configuration are marked by the red dots in \textbf{(a)}.}
\label{Fig5}
\end{figure}

To give further insights, say as an example we are "loading" with $\hat{\gamma}_{1}>0$ fixed; see Fig. $\ref{Fig5}$. Intuitively, we might say that since we reach $\hat{\lambda}_{1}=\sqrt{3}\hat{\lambda}_{1}^{L}$ first, the initial bifurcation is into a kink-wave solution. However, a direct transition from the primary axial tension state to a fully developed kink-wave solution is only possible if the perturbations applied to the tube are large in amplitude \cite{ericksen1975}. Given that our "loading" is extremely controlled, such a bifurcation is not physically feasible. Thus, the initial bifurcation is as generally expected into a localised necking solution as $\hat{\lambda}_{1}\rightarrow\hat{\lambda}_{1}^{L}$ (i.e. as $\lambda\rightarrow\lambda_{\text{cr}}^{L}$), and this occurs sub-critically since $\hat{\lambda}_{1}<\hat{\lambda}_{1}^{L}$ must hold by $(\ref{lam1int})$; see Fig. $\ref{Fig5}$ (a).
Beyond this point, we expect that this necking solution evolves into a kink-wave solution. In other words, the neck or depression will first undergo a propagation in the radial direction to a near maximum amplitude followed by an axial propagation; see Fig. $\ref{Fig5}$ (b). The final "two-phase" or "kink-wave" configuration consists of a thin section with stretch $\lambda_{R}$ centred at $z=0$ in between two thick sections with stretch $\lambda_{L}$. These thick and thin sections are connected by a smooth transition region whose length is assumed to be negligible, and the overall average stretch of the tube is $\lambda_{\text{cr}}^{L}$ plus some small increment. Since we are "loading", the stretch as $z\rightarrow\pm\infty$ is also $\lambda_{\text{cr}}^{L}$ plus some small increment near the initial bifurcation point, which is why the thicker sections are on the outside. However, we note that whether or not the transition from necking to kink-wave solution is sudden or gradual is largely dependent on the stability of the solution $(\ref{ampsollim})$ which we do not analyse here. If we were "unloading",  the initial bifurcation would of course be into a localised bulge as $\hat{\lambda}_{1}\rightarrow\hat{\lambda}_{1}^{R}$, and the thicker section of the ensuing kink wave solution would be centred at $z=0$ instead.
\subsection{Taking $\gamma$ as the control parameter with $\lambda$ fixed \label{sec4b}}
An alternate approach is to subject the tube to a fixed axial stretch $\lambda$ initially and then increase the surface tension monotonically from zero. Varying the surface tension can be achieved chemically by lowering the shear modulus of the material through temperature change, say. We set
\begin{align}
\gamma&=\gamma_{\tx{cr}}+\varepsilon\,\tilde{\gamma}_{1},
\end{align}
where $\tilde{\gamma}_{1}$ is a constant of $O(1)$ and the bifurcation values $\gamma_{\tx{cr}}$ satisfy $(\ref{bifconn})$ with $\lambda_{\tx{cr}}$ replaced by a fixed $\lambda$ on the right-hand side. Then, by applying the same solution procedure as presented in the previous section, we obtain the amplitude equation
\begin{align}
\scr{A}''&=\tilde{\gamma}_{1}\,\tilde{\kappa}_{1}\,\scr{A}+\tilde{\kappa}_{2}\,\scr{A}^{2}, \label{AEg}
\end{align}
whose solitary wave solution is $\scr{A}=-\{(3\tilde{\gamma}_{1}\tilde{\kappa}_{1})/(2\tilde{\kappa}_{2})\}\,\text{sech}^{2}(\sqrt{\tilde{\gamma}_{1}\tilde{\kappa}_{1}}s/2)$. Unlike its counterpart $\kappa_{1}$ in the previous loading scenario, $\tilde{\kappa}_{1}$ is generally negative, and this can be explained as follows. In this loading scenario, the non-trivial eigenvalues are $\pm\alpha_{1}=\pm\sqrt{\tilde{\gamma}_{1}\,\tilde{\kappa}_{1}}$. Since for \textit{any} fixed $\lambda$ the constant $\tilde{\gamma}_{1}$ is negative (resp. positive) in the stable (resp. unstable) regime, we require that $\tilde{\kappa}_{1}<0$ for the expected exponential (resp. periodic) behaviour to occur. It follows that the bifurcation is sub-critical since the term $\sqrt{\tilde{\gamma}_{1}\tilde{\kappa}_{1}}$ in the solitary wave solution forces $\tilde{\gamma}_{1}$ to be negative. Based on the findings for a solid cylinder \cite{FuST}, we expect that the relation $\tilde{\kappa}_{2}=\lambda^{2}\,\tilde{\kappa}_{1}\,d\gamma_{\text{cr}}/d\lambda$ holds true, and we have verified this numerically for the Gent material model and explicitly for the neo-Hookean model. The localised solution to $(\ref{AEg})$ is therefore necking for $\tilde{\kappa}_{2}>0$ and bulging for $\tilde{\kappa}_{2}<0$.
\begin{figure}[h!]
\centering
\begin{subfigure}[t]{0.483\textwidth}
\vspace{1mm}
\includegraphics[width=\linewidth, valign=t]{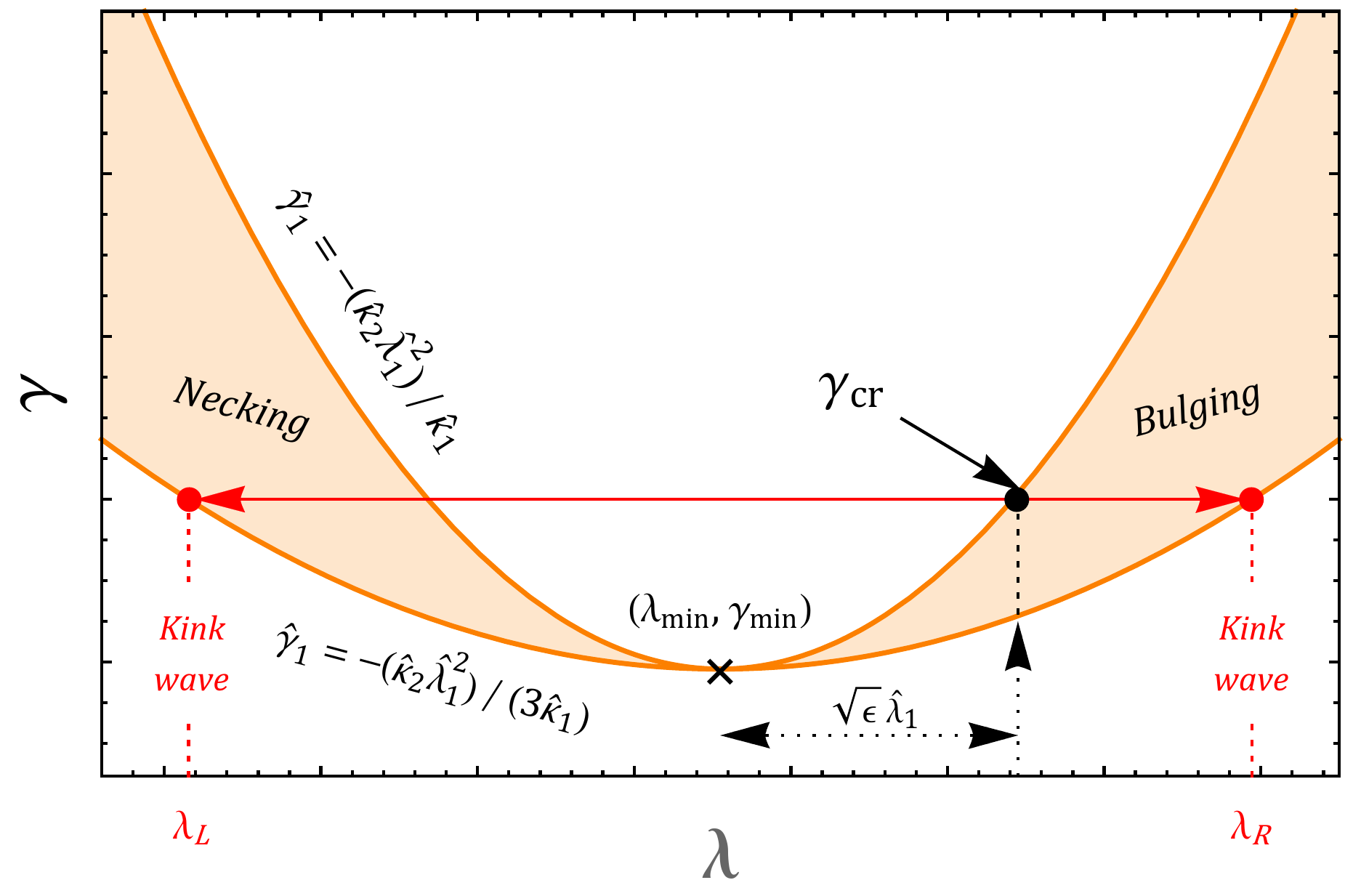}
\subcaption*{\textbf{(a)}}
\end{subfigure}\hfill
\begin{subfigure}[t]{0.483\textwidth}
\includegraphics[width=\linewidth, valign=t]{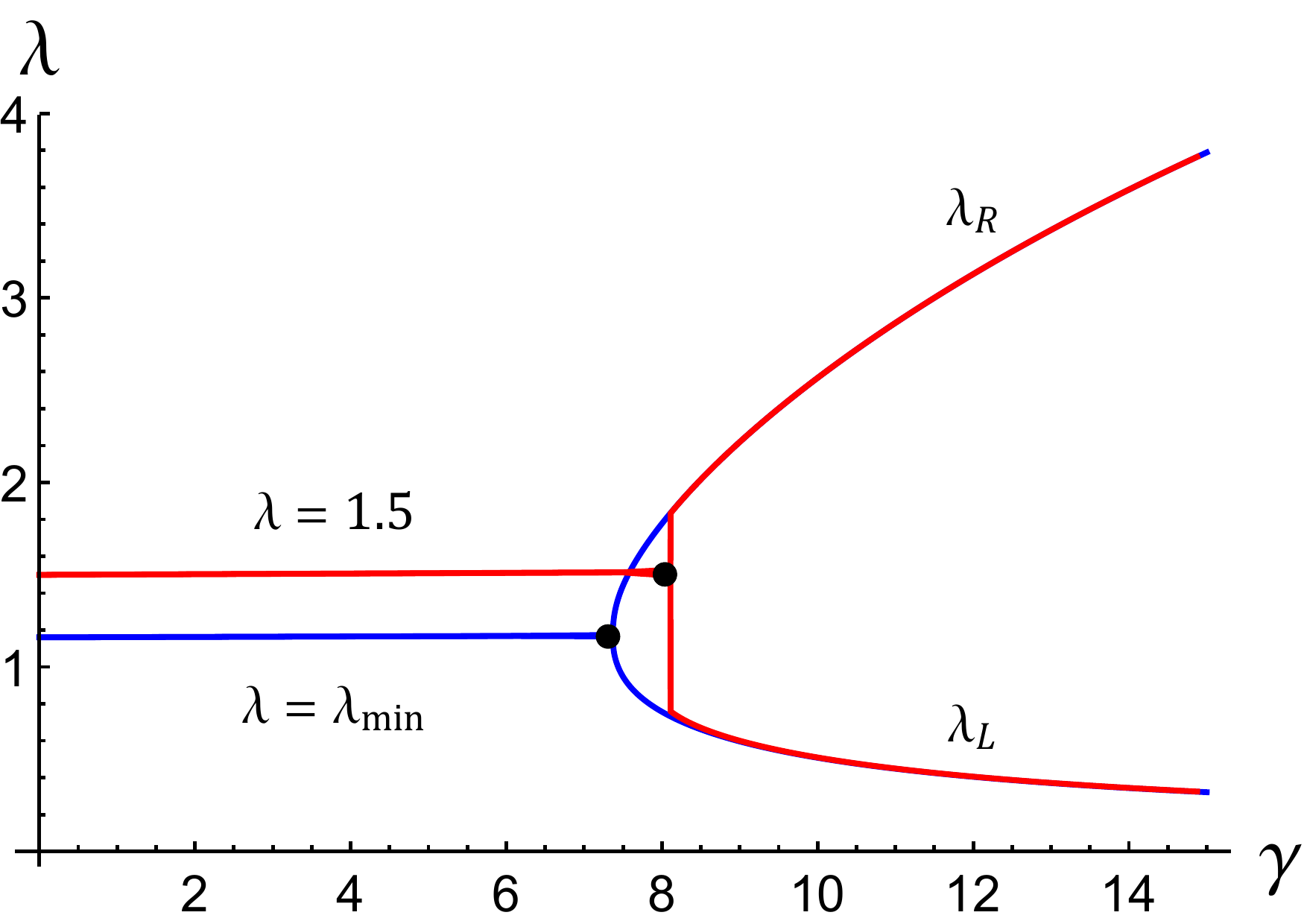}
\vspace{1mm}
\subcaption*{\textbf{(b)}}
\end{subfigure}
\caption{\textbf{(a)} If we fix $\hat{\lambda}_{1}>0$ and then increase $\gamma$ gradually from zero, the initial bifurcation is into a localised bulge, and this occurs sub-critically in the limit $\hat{\gamma}_{1}\rightarrow -\hat{\kappa}_{2}\hat{\lambda}_{1}^{2}/\hat{\kappa}_{1}$ (as marked by the black dot). A "snap-through" to the two red dots at $\lambda=\lambda_{L,R}$ then occurs (i.e. the localised bulge "jumps" to a kink-wave solution). \textbf{(b)} FEM simulations from \cite{EmeryFu2020} which verify the "jump" behaviour for fixed $\lambda\neq\lambda_{\text{min}}$. The black dots mark the bifurcation points given by $(\ref{bifconn})$. Exceptionally, no "jump" occurs when $\lambda=\lambda_{\text{min}}$, and a continuous, supercritical transition to a kink-wave solution occurs instead (blue curve).}
\label{Fig6}
\end{figure}

In the limit $\lambda\rightarrow\lambda_{\tx{min}}$, $\tilde{\kappa}_{2}$ vanishes and the solitary wave solution to $(\ref{AEg})$ diverges. Through appropriate re-scaling of $\scr{A}$, the amplitude solution $(\ref{ampsollim})$ is found to be valid in this limiting case also for $-\hat{\kappa}_{2}\hat{\lambda}_{1}^{2}/(3\hat{\kappa}_{1})<\hat{\gamma}_{1}< -\hat{\kappa}_{2}\hat{\lambda}_{1}^{2}/\hat{\kappa}_{1}$. To illustrate further, say we fix $\hat{\lambda}_{1}>0$ and increase $\gamma$ gradually from zero; see Fig. $\ref{Fig6}$ (a). Then, with the aid of $(\ref{lim1})$ and $(\ref{adf1})-(\ref{adf2})$, we deduce that the initial bifurcation occurs sub-critically in the limit $\hat{\gamma}_{1}\rightarrow -\hat{\kappa}_{2}\hat{\lambda}_{1}^{2}/\hat{\kappa}_{1}$ (i.e. as $\gamma\rightarrow\gamma_{\text{cr}}$), and takes the explicit form of a localised bulge (since $\hat{\lambda}_{1}>0$). Based on previous FEM simulations \cite{EmeryFu2020}, we expect that a "snap-through" to the two red dots at $\lambda=\lambda_{L,R}=\lambda_{\text{min}}\mp\sqrt{3\varepsilon}\hat{\lambda}_{1}$ will then occur; see Fig. $\ref{Fig6}$ (a). In other words, a "snap-through" from localised bulging to a kink-wave solution will take place. For the latter, the configuration consists of a thicker section with stretch $\lambda_{L}$ centred at $z=0$ in between two thinner sections with stretch $\lambda_{R}$, with the overall average stretch remaining fixed at $\lambda=\lambda_{\text{min}}+\sqrt{\varepsilon}\hat{\lambda}_{1}$.

In the special case $\lambda=\lambda_{\tx{min}}$, the solution $(\ref{ampsollim})$ cannot capture the bifurcation behaviour since the inequalities $(\ref{NCond})$ cannot be satisfied at $\hat{\lambda}_{1}=0$.  However, we may make the substitution $\hat{\scr{A}}(\hat{s})= 2^{-1} \lambda_{\tx{min}}^{-2} \sqrt{- 3\,\hat{\gamma}_{1}\hat{\kappa}_{1}/\hat{\kappa}_{2}}  \, h(t) $, with $t=\sqrt{-\hat{\gamma}_{1}\hat{\kappa}_{1}} \hat{s}$, to reduce the amplitude equation \rr{AEl} to
$
h''(t)= h(h^{2}-1) 
$ in this limit.
The latter equation has previously been derived by Xuan and Biggins \cite{xuan2017}, and it admits the kink-wave solution  $h(t)={\rm tanh} (t /\sqrt{2})$ that tends to $\pm 1$ as $t \to \pm \infty$, respectively.
As we increase $\gamma$ beyond the associated bifurcation value $\gamma_{\tx{min}}$, a \textit{continuous} transition to this kink-wave solution occurs, and the bifurcation is exceptionally super-critical as shown by the blue curve in Fig. $\ref{Fig6}$ (b).

\section{Further insights into Case 3 \label{sec6}}
In this section, we provide further insights into the bifurcation behaviours in Case $3$. As an illustrative example, we fix $\lambda>1-A^2$ and take $\gamma$ as the control parameter. For the sake of brevity, the neo-Hookean strain-energy is deployed here, but an extension to the Gent model can be achieved as shown previously. By applying the same approach as in section $\ref{sec4}$ $\ref{sec4b}$, the corresponding amplitude equation is determined to be of the form $(\ref{AEg})$. The counterpart of $\tilde{\kappa}_{1}$ in Case 3 is still generally negative, and $\tilde{\kappa}_{2}=\lambda^{2}\tilde{\kappa}_{1}\,d\gamma_{\tx{cr}}/d\lambda$.

\begin{figure}[h!]
\centering
\begin{subfigure}[t]{0.5\textwidth}
\includegraphics[width=\linewidth, valign=t]{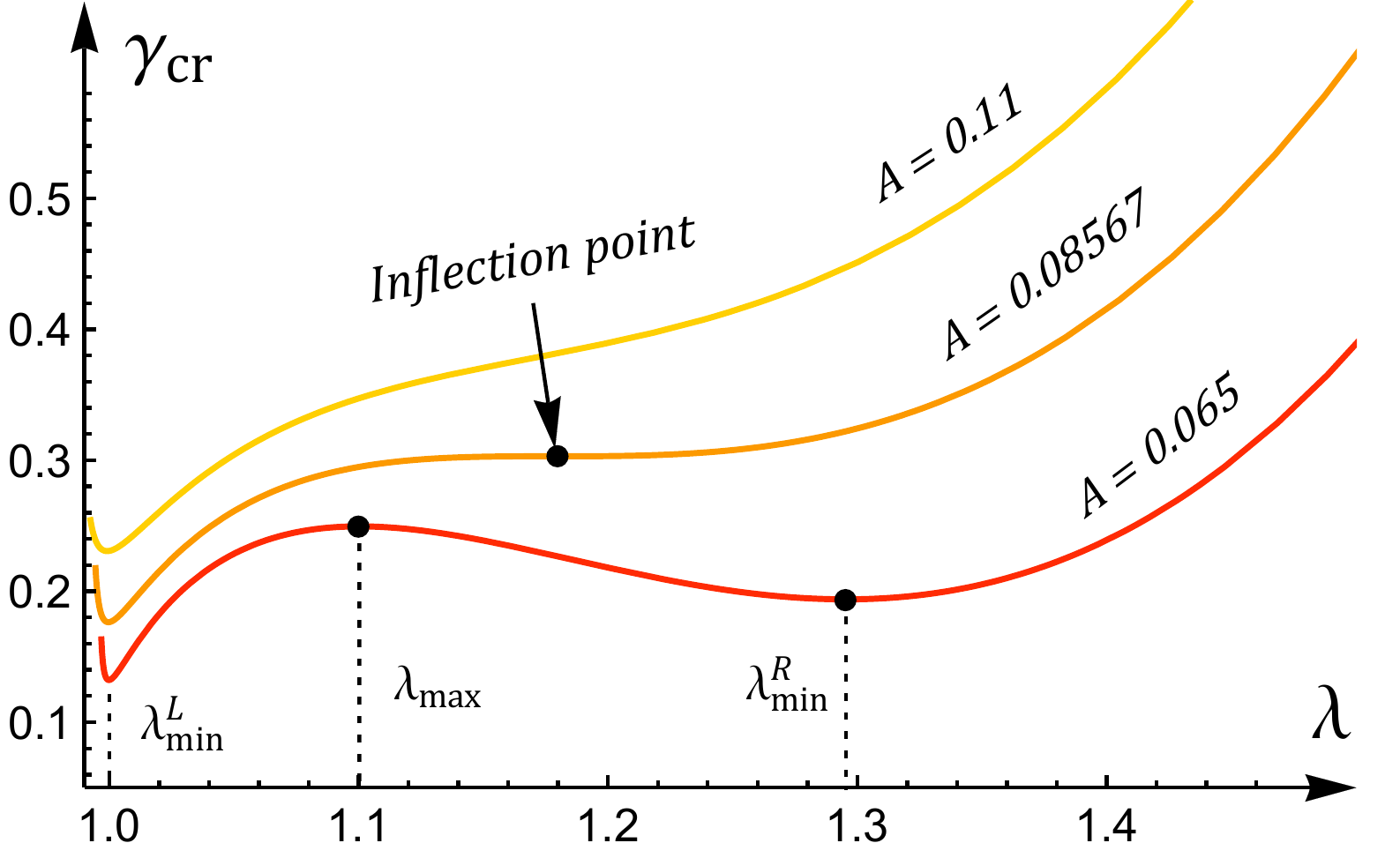}
\subcaption*{\textbf{(a)}}
\end{subfigure}\hfill
\begin{subfigure}[t]{0.48\textwidth}
\includegraphics[width=\linewidth, valign=t]{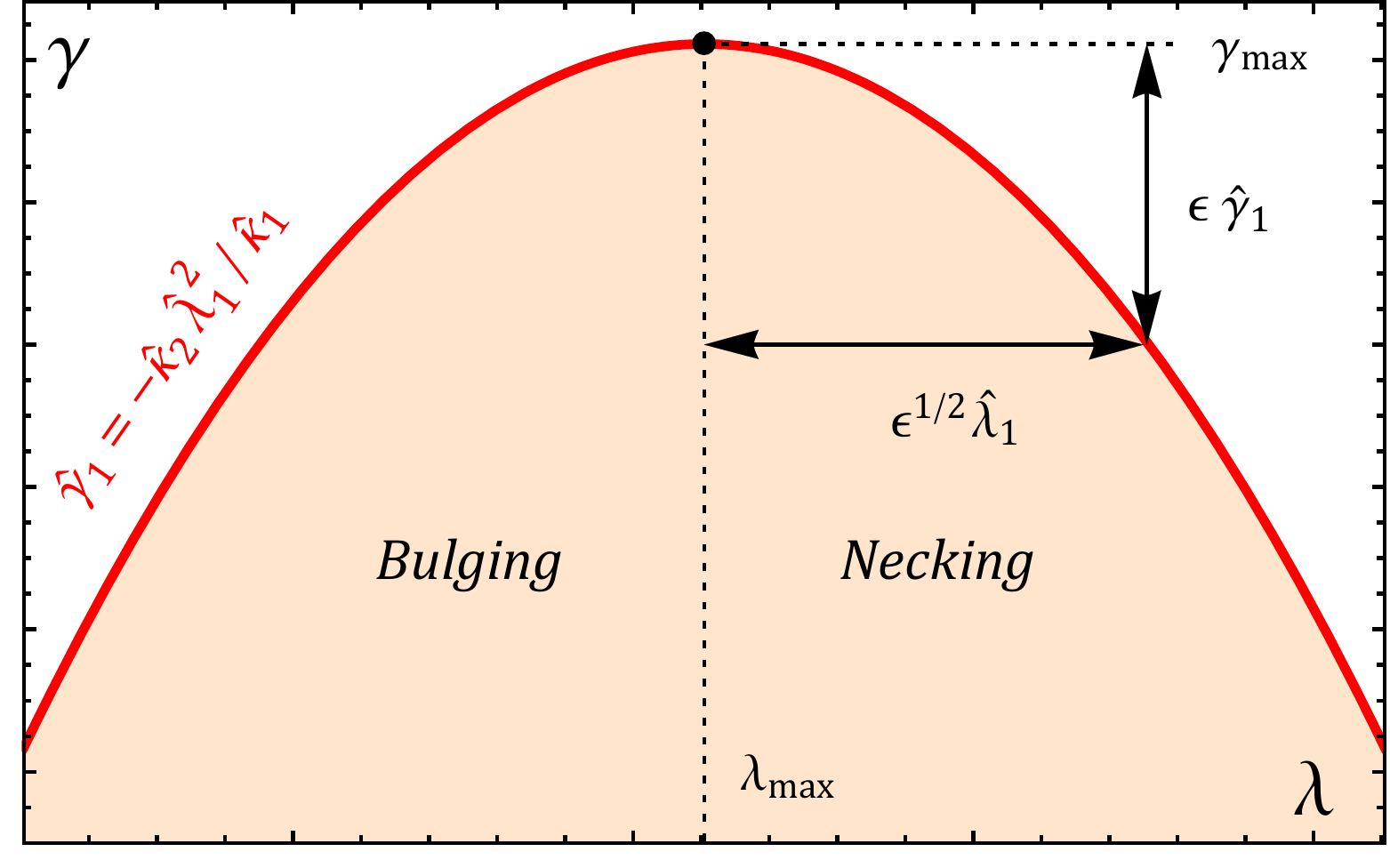}
\vspace{1mm}
\subcaption*{\textbf{(b)}}
\end{subfigure}
\caption{\textbf{(a)} Plots of $\gamma_{\tx{cr}}$ against $\lambda$ for $A=0.065$ (red), $A=0.08567$ (orange) and $A=0.11$ (yellow). On the curve corresponding to $A=0.065$, the black dots mark the two additional local extrema of $\gamma_{\tx{cr}}$ with respect to $\lambda$ which emerge in the large thickness regime. In the limit $A\rightarrow 0.08567^{-}$, these two extrema coalesce to form an inflection point (as shown by the arrow), and above this threshold, $\gamma_{\tx{cr}}$ has a single minimum value. \textbf{(b)} A blow up of $(\ref{bifconn3})$ in the large thickness regime about $\lambda=\lambda_{\tx{max}}$. The red curve gives the bifurcation points $\hat{\gamma}_{1}=-\hat{\kappa}_{2}\hat{\lambda}_{1}^{2}/\hat{\kappa}_{1}$ in a small neighbourhood of $\gamma_{\tx{max}}$, whilst the orange region represents $\hat{\gamma}_{1}<\hat{\kappa}_{2}\hat{\lambda}_{1}^{2}/\hat{\kappa}_{1}$, the domain of existence of localisation in this limiting case.}
\label{Fig7}
\end{figure}

Whilst the structure of the amplitude equation appears near identical to the Case 2 counterpart, there is a subtlety which alters the interpretation of the bifurcation solutions somewhat, and this can be explained as follows. In Case 2, $\gamma_{\tx{cr}}$ as a function of $\lambda$ has a \textit{single} minimum for any tube thickness, as was illustrated in Fig. $\ref{Fig2}$. In Case 3 however, there exists a threshold value for $A$ below which two \textit{additional} local extrema (i.e. a maximum and a minimum) of $\gamma_{\tx{cr}}$ with respect to $\lambda$ emerge. For the neo-Hookean strain energy, this threshold is $A\approx 0.08567$, and in the limit $A\rightarrow0.08567^{-}$ these two additional extrema of $\gamma_{\tx{cr}}$ coalesce to form an inflection point; see Fig. $\ref{Fig7}$ (a). In the limit $A\rightarrow 0$, incompressibility may only be satisfied for $\lambda=1$, and the associated bifurcation point $\gamma_{\tx{cr}}/A \rightarrow 2$ corresponds to localisation in a cylindrical cavity surrounded by an infinite solid \cite{henann,xuan2016}. For values of $A$ above 0.08567, the solitary wave solution corresponds to a localised neck for $1-A^{2}<\lambda<\lambda_{\tx{min}}$ and a localised bulge for $\lambda>\lambda_{\tx{min}}$.

For $A<0.08567$, we denote by $\lambda_{\tx{max}}$ and $\lambda^{L,R}_{\tx{min}}$ the values of $\lambda$ at the maximum and the left and right hand minima of $\gamma_{\tx{cr}}$. Then, given the form of $\tilde{\kappa}_{2}$, it follows that the localised solution in this large thickness regime corresponds to necking for $1-A^{2}<\lambda<\lambda_{\tx{min}}^{L}$ and $\lambda_{\tx{max}}<\lambda<\lambda^{R}_{\tx{min}}$, and bulging for $\lambda_{\tx{min}}^{L}<\lambda<\lambda_{\tx{max}}$ and $\lambda>\lambda^{R}_{\tx{min}}$.
Now, in the limit $\lambda\rightarrow\lambda^{L,R}_{\tx{min}}$, we find that a Case 3 counterpart of the localised solution $(\ref{ampsollim})$ and the associated post-bifurcation behaviour is valid. However, in the limit $\lambda\rightarrow\lambda_{\tx{max}}$, a key difference lies in the fact that $\hat{\kappa}_{2}$ is no longer positive; see $(\ref{adf2})$. This means that, given the expansions $\gamma=\gamma_{\tx{max}}+\varepsilon\,\hat{\gamma}_{1}$ and $\lambda=\lambda_{\tx{max}}+\varepsilon^{1/2}\hat{\lambda}_{1}$, and the form of $\hat{\scr{A}}_{0}^{\pm}$ in $(\ref{GSs})$, the inequality sign in the necessary condition $(\ref{NCond})_{2}$ for localisation is flipped.  Therefore, in the vicinity of $\lambda_{\tx{max}}$, the localised solution $(\ref{ampsollim})$ instead exists for $\hat{\gamma}_{1}<-\hat{\kappa}_{2}\hat{\lambda}_{1}^{2}/\hat{\kappa}_{1}<0$, and thus occurs sub-critically; see Fig. $\ref{Fig7}$ (b). Also, on flipping the inequality sign in $(\ref{NCond})_{2}$, we see that $(\ref{NCond})$ can be satisfied at $\hat{\lambda}_{1}=0$. Thus, the Case 3 counterpart of $(\ref{ampsollim})$ for $A<0.08567$ is valid at $\lambda=\lambda_{\text{max}}$, and reduces to the dark solitary wave $\hat{\scr{A}}=-\lambda_{\text{max}}^{-2}\sqrt{(-3\hat{\gamma}_{1}\hat{\kappa}_{1})/(2\hat{\kappa}_{2})}\,\text{sech}(\sqrt{\hat{\gamma}_{1}\hat{\kappa}_{1}}\hat{s})$.
\section{Conclusion \label{sec7}}
If one lateral surface of a soft slender tube is in smooth contact with a rigid boundary such that radial displacement and surface tension are prohibited, and the other lateral surface remains traction-free and under surface tension, localised bifurcation solutions become widely favourable over periodic axisymmetric modes. Contrary to our original thoughts, this statement is also true when both lateral surfaces are traction-free, though in this case these localised solutions are preceded by elliptic circumferential buckling modes \cite{emery2021elasto}. Throughout this work, we have referred to the latter scenario as Case 1, whilst Cases 2 and 3 pertain to the scenarios where the inner or outer surfaces, respectively, are radially fixed.

Through a weakly non-linear analysis formulated in terms of a general material model, we have explicitly characterised the localised solutions which initially bifurcate from the primary state of axial tension in Case 2. These theoretical results pertain to two common scenarios of loading: a fixed surface tension with monotonically varying axial stretch, or a fixed averaged axial stretch with monotonically increasing surface tension. In the former scenario, the fixed surface tension $\gamma$ attains a minimum $\gamma_{\tx{min}}$ at the critical axial stretch $\lambda_{\tx{cr}}=\lambda_{\tx{min}}$, and there consequently exists two bifurcation values $\lambda_{\tx{cr}}^{L}<\lambda_{\tx{min}}$ and $\lambda_{\tx{cr}}^{R}>\lambda_{\tx{min}}$ for any fixed $\gamma>\gamma_{\tx{min}}$. We explained how, on assuming that the resultant axial force $\mathcal{N}$ is zero initially and then increasing $\lambda$ ("loading"), the left bifurcation point $\lambda_{\tx{cr}}^{L}$ is always encountered first, and the associated \textit{sub-critical} bifurcation solution is localised necking. In contrast, when applying a dead load to an end of the tube initially and decreasing $\lambda$ ("unloading"), the right bifurcation point $\lambda_{\text{cr}}^{R}$ is encountered first and the associated localised solution is bulging. When fixing $\lambda$ and increasing $\gamma$, the localised solution was found to be necking for $\lambda<\lambda_{\tx{min}}$ and bulging for $\lambda>\lambda_{\tx{min}}$. In Case 3, two additional local extrema of the critical surface tension emerge in the regime of large tube thickness $A<0.08567$, meaning that necking and bulging can occur in several separate intervals of axial stretch.

In both loading scenarios, an appropriate rescaling reveals the existence of a small region either side of $\lambda=\lambda_{\tx{min}}$ wherein a variety of localised bifurcation behaviours exist. Based on these theoretical results, we provided a complete interpretation of the expected post-bifurcation behaviour which is supported by FEM simulations \cite{EmeryFu2020}. For fixed $\lambda$, we explained how the tube will initially admit a localised necking or bulging solution which "jumps" to a final kink-wave solution consisting of a bulged and depressed region with constant but distinct axial stretches $\lambda_{L}$ and $\lambda_{R}$, respectively. These stretches are given analytically in the vicinity of $\lambda_{\tx{min}}$. At $\lambda=\lambda_{\tx{min}}$, the bifurcation is exceptionally supercritical, and the corresponding kink wave solution has been derived explicitly.

We conclude by noting that, as a first attempt, we have used the simplest liquid-like surface model to describe surface effect. However, more sophisticated models that take area stretch into account \cite{gurtin1975,huang2006}, or even surface stiffness into account \cite{steigmann1997}, may be used.
\vskip6pt

\enlargethispage{20pt}

\dataccess{This article has no additional data.}

\aucontribute{\textbf{Dominic Emery and Yibin Fu:} Conceptualisation, Methodology, Software, Writing - original draft, Writing review and editing.}

\competing{The authors declare they have no competing interests.}


\ack{We would like to thank the referees of this paper for their constructive and knowledgeable comments and suggestions. The first author (DE) also acknowledges the School of Computing and Mathematics, Keele University for supporting his PhD studies through a faculty scholarship.}

\appendix

\bibliographystyle{RS} 

\end{document}